\documentclass[aps,prx,twocolumn,groupedaddress]{revtex4-2}
\usepackage{latexsym}
\usepackage{amsmath}
\usepackage{amsfonts}
\usepackage{amssymb}
\usepackage{amscd}
\usepackage{bbm}
\usepackage{bbold}
\usepackage{tikz}
\usepackage{tikz-cd}
\usetikzlibrary{shapes}
\usetikzlibrary{patterns,snakes}
\usepackage{hyperref}
\usepackage[export]{adjustbox}
\usepackage[caption=false]{subfig}
\usepackage{soul}

\newcommand{\Z}{\mathcal{Z}}
\newcommand{\ket}[1]{|#1\rangle}
\newcommand{\bra}[1]{\langle#1|}
\newcommand{\bracket}[2]{\langle#1|#2\rangle}
\newcommand{\vev}[1]{\langle#1\rangle}

\begin{document}

\title{Efficient simulations of epidemic models with tensor networks: application to the one-dimensional SIS model} 

\author{Wout Merbis}
\email[]{w.merbis@uva.nl}
\author{Cl\'elia de Mulatier}
\author{Philippe Corboz}

\affiliation{Dutch Institute for Emergent Phenomena (DIEP) \& Institute for Theoretical Physics (ITFA), University of Amsterdam, 1090 GL Amsterdam, The Netherlands}

\date{\today}

\begin{abstract}
The contact process is an emblematic model of a non-equilibrium system, containing a phase transition between inactive and active dynamical regimes. In the epidemiological context, the model is known as the susceptible-infected-susceptible (SIS) model, and widely used to describe contagious spreading. In this work, we demonstrate how accurate and efficient representations of the full probability distribution over all configurations of the contact process on a one-dimensional chain can be obtained by means of Matrix Product States (MPS). We modify and adapt MPS methods from many-body quantum systems to study the classical distributions of the driven contact process at late times. We give accurate and efficient results for the distribution of large gaps, and illustrate the advantage of our methods over Monte Carlo simulations. Furthermore, we study the large deviation statistics of the dynamical activity, defined as the total number of configuration changes along a trajectory, and investigate quantum-inspired entropic measures, based on the second R\'enyi entropy. 
\end{abstract}

\maketitle

\section{Introduction}

Accurate and efficient mathematical modeling of complex systems, composed out of many interacting constituents, poses a formidable challenge in many biophysical and socio-ecological systems \cite{lloyd2001viruses,albert2002statistical,boccaletti2006complex,castellano2009statistical,vespignani2012modelling,artime2022origin}. Due to random nature of biochemical, social or ecological interactions, models are often defined stochastically in terms of discrete state Markov chains, describing the continuous time evolution of a probability vector over the set of all possible systems states \cite{gleeson2013binary,masuda2017random,ghavasieh2020statistical,merbis2023emergent}. As the number of system states grows exponentially with system size, exact methods quickly become intractable and one is often forced to rely on Monte Carlo simulation \cite{privman1990finite} or mean-field approximations 
\cite{gleeson2012accuracy,gao2016universal,kiss2017mathematics}. 
In the mean field approximation, the focus is often to compute average quantities, disregarding variances around these quantities. This can lead to problems in the context where systems are non-self averaging, which is often the case for processes with reproduction and annihilation, such as branching processes in epidemic modeling, population dynamics \cite{young2001reproductive,houchmandzadeh2009theory}, or nuclear reactor physics \cite{dumonteil2014particle}. 
A concrete and urgent application where this is relevant, concerns the modeling and forecasting of contagion spreading \cite{hethcote2000mathematics,pastor2001epidemic,sahneh2013generalized,PastorSatorras2015,merbis2023complex}, imperative in light of the recent COVID-19 pandemic. 

The problem of finding accurate and efficient descriptions of high-dimensional vectors is a familiar one in the study of emergent collective behavior in composite quantum systems. Many-body quantum systems are characterized by an exponentially large Hilbert space, although this is often merely a facade \cite{poulin2011quantum}. The subset of physically relevant states is often much smaller (and scaling polynomially in system size), allowing for efficient representation using tensor networks (TN) \cite{white1992density,schollwock2011density,orus2014practical}. TN exploit an internal structure of high-dimensional composite wavefunctions, by representing them as a set of lower dimensional tensors, contracted along internal (bond) indices. The bond dimension of the tensor network represents the size of the effective state space of the system, which is often much smaller than the full Hilbert space. Efficient representations, where the bond dimensions scales at most polynomially with system size, are usually possible when interactions between the constituents are local. 

In this regard, stochastic models of composite complex systems display similarities with quantum many-body systems \cite{schonberg1952application,doi1976second,grassberger1980fock,peliti1985path,schutz1995reaction,hinrichsen2000non,SCHUTZ20011,merbis2022logistic}. The high-dimensional probability vector is an element of the tensor product space of the probability vectors of the constituents. A natural question is whether and when these high-dimensional probability vectors can be efficiently represented using TN. In other words, is it possible to use TN to find an optimal compression of large-dimensional probability distributions, while remaining as close as possible to the original distribution? 

TN methods are known to be effective for stochastic kinetic models such as totally asymmetric simple exclusion process (TASEP) 
\cite{derrida1998exact,Johnson2010,de2011large,lazarescu2011exact,gorissen2012exact}. Methods based on Density Matrix Renormalization Group (DMRG) have been applied to stochastic out-of-equilibrium systems in \cite{nishino1995density,carlon1999density,gorissen2009density,hooyberghs2010thermodynamics}. More recently, methods based on Projected Entangled Pair States (PEPS) have been applied to stochastic models of non-equilibrium systems \cite{helms2020dynamical,causer2022optimal} and Matrix Product States (MPS) have been used for studying large deviation statistics of dynamical stochastic systems \cite{helms2019dynamical,banuls2019using,causer2022finite,strand2022using,garrahan2022topological}. 
Additionally, using TN provides certain advantages over Monte Carlo (MC) methods, frequently used for analyzing non-equilibrium stochastic systems. Instead of computing empirical averages over a large number of numerically generated trajectories, the tensor network is capable of directly providing accurate information on observables averaged over the ensemble of all dynamical trajectories. It is therefore perfectly suited to provide efficient computations when variances are large \cite{Johnson2014}, and give accurate predictions on probability distributions of rare events \cite{causer2021optimal}.

The majority of work in this direction is based on stochastic systems satisfying local detailed balance, an assumption which does not necessarily hold for all relevant biophysical systems, and certainly not for most epidemic models.
In this paper, we investigate the use of TN for epidemiological modeling by constructing an MPS representation for the one-dimensional contact process \cite{harris1974contact,grassberger1983critical,jensen1993time,hooyberghs2001one}, known in the epidemiological literature as the Susceptible-Infected-Susceptible (SIS) model \cite{hethcote2000mathematics,PastorSatorras2015}. The contact process contains sites which may be occupied (infected) or empty (susceptible) and allows transitions from the empty to the occupied state only if a neighboring site is occupied (infection). In turn, occupied sites may transition to empty regardless of their neighbors state, modeling the recovery of individuals. The process has been called `the Ising model of absorbing state transitions', and serves as a natural starting point for developing new methods for non-equilibrium problems \cite{marro2005nonequilibrium}. 
The special aspect of this work lies in the fact that the contact process does not satisfy local detailed balance, such that there is no map between the Markov generator and a Hermitian matrix. Hence, known MPS algorithms used for Hermitian time-evolution have to be adapted for the purely stochastic evolution of the contact process, with applications to more generic biophysical and socio-ecological models.

The power of the MPS representation lies in providing an efficient representation for the $2^N$ dimensional probability vector of the model, by finding a lower dimensional representation which keeps only the relevant physical states.
This makes it possible to efficiently compute marginals, conditional probabilities, expectation values and higher moments for any macroscopic observable one might be interested in. There is hence great potential for TN as a modeling and forecasting tool, which will be able to provide more detailed information on correlations and variances than conventional methods, without having to rely on costly sampling-based simulations.
To illustrate the advantage of TN as a numerical tool complementary to MC simulations, we use the MPS representation for the one-dimensional contact process to compute the expectation value for finding large gaps in the occupancy in the chain. Similar distributions of gaps has been studied in the context of critical branching processes \cite{ramola2014universal}.
These large gaps (between successive infected sites) constitute extremely rare events in both the absorbing and active phases of the dynamical process, and are hence difficult to obtain with MC simulation. 

We additionally study the thermodynamics of dynamical trajectories \cite{lecomte2007thermodynamic,garrahan2009first,chetrite2015nonequilibrium} and the large deviation theory \cite{touchette2009large} for the dynamical activity in the 1d contact process using the MPS. The moment generating function for the dynamical activity (defined as the number of configuration changes per unit time) with dual parameter $s$, can be obtained as the norm of a tilted probability vector, which evolves in time by a tilted generator \cite{lecomte2007thermodynamic,hooyberghs2010thermodynamics}. This is obtained from the Markov generator by exponentially tilting the off-diagonal terms by $e^s$, such that for $s>0$ active trajectories become more likely, while $s<0$ tilts the dynamics towards inactive trajectories. The Scaled Cumulant Generating Function (SCGF) for the dynamical activity at late times can then be obtained as the leading eigenvalue of the tilted generator \cite{garrahan2009first}, which we obtain using variational MPS methods. We find this leads to a clear kink in the SCGF, indicating a discontinuous phase transition between the active and inactive phases as a function of control parameter $s$, in accordance with previous studies using DMRG methods \cite{hooyberghs2010thermodynamics}. 

The MPS representation invites one to study quantum information inspired measures for classical systems.
Although classical information measures, such as the full Shannon entropy and mutual information between two subsets of the chain, remain hard to compute on the space of all state configurations, one can efficiently compute the second R\'enyi entropy \cite{wolf2008area,schuch2008entropy,scalet2021computable}. We investigate the second R\'enyi entropy, the R\'enyi entranglement entropy and R\'enyi Mutual Information (RMI) for the one-dimensional contact process as a function of transmission rate $\lambda$ and dual parameter $s$. We find that, in accordance to phase transitions in equilibrium models \cite{iaconis2013detecting}, the RMI can be used to detect the absorbing state transition. This demonstrates that the RMI also provides a way to detect non-equilibrium phase transitions independent of a choice of order parameter.

This work is organized as follows, in section \ref{sec:model} we introduce the model and discuss different driving protocols which remove the absorbing state from the system, such that the process is driven to a Non-Equilibrium Steady State (NESS) at late times. Section \ref{sec:MPSrep} introduces the Matrix Product State used to approximate the complete $2^N$ dimensional probability distribution over all configurations of the chain. In section \ref{sec:thermo} we review the thermodynamics of trajectories and discuss the large deviation principle. Section \ref{sec:varNESS} deals with details on the variational MPS method, its accuracy and efficiency and here we construct the MPS representation of the NESS to study the distribution of gaps in the chain. Results on the SCGF for the dynamical activity are presented in section \ref{sec:varMPS}. Section \ref{sec:entropic} deals with the entropic measures based on (second) R\'enyi entropies obtained from the MPS representation. We comment in the discussion section \ref{sec:conclusions} on possible future work towards TN as a new modeling paradigm for epidemic spreading phenomena, or other non-equilibrium statistical models of complex systems. 



\section{The model}\label{sec:model}

\begin{figure*}[!ht]
    \centering
    \includegraphics[width = 0.8\textwidth]{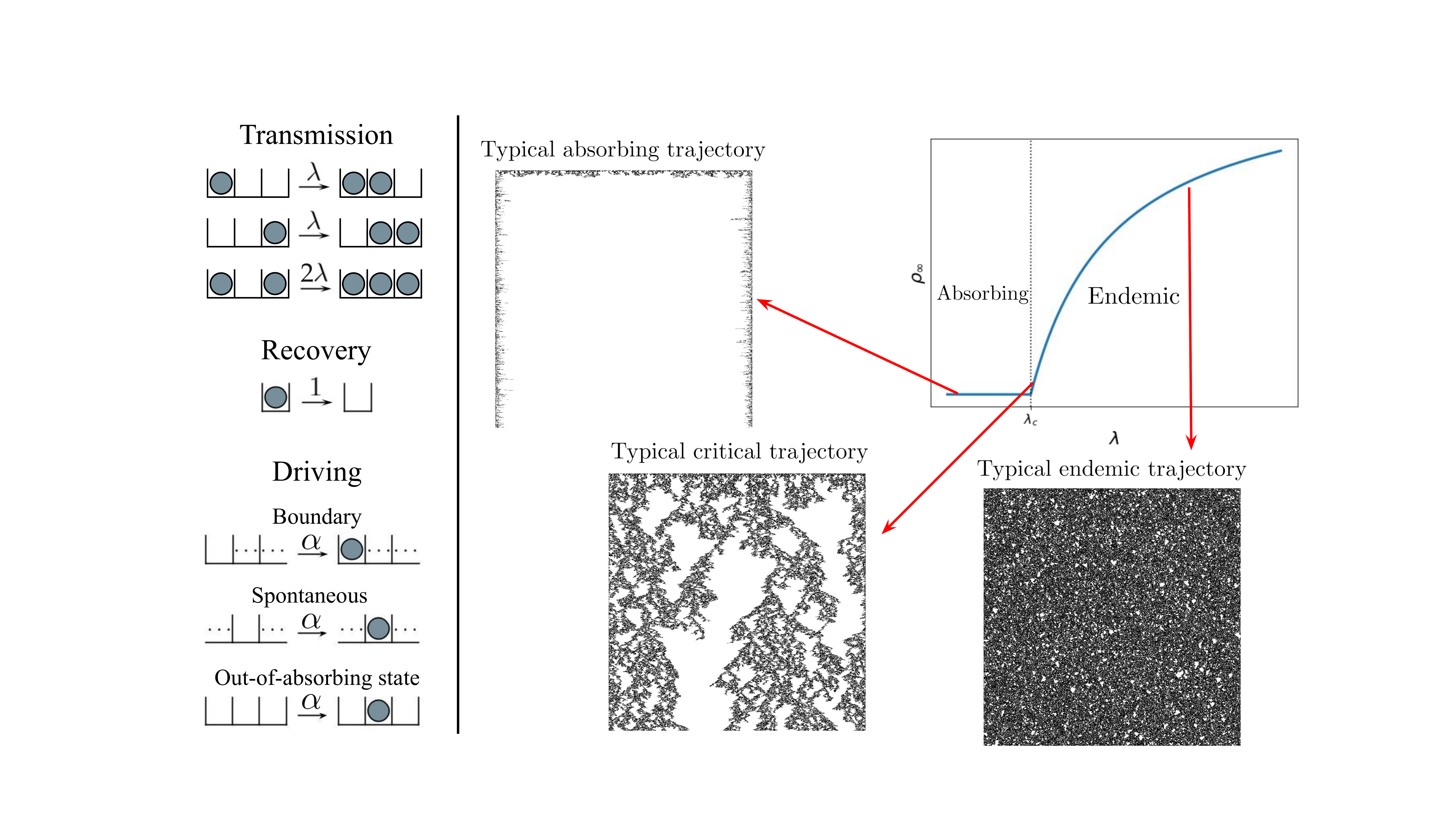}
    \caption{
    Left: The possible transitions for the contact process are transmission (each infected site can infect a susceptible neighbor with rate $\lambda$) and recovery (infected sites become susceptible with rate 1). 
    Additionally, to maintain the process out of equilibrium, we consider three possible driving terms: the spontaneous infection of a site either at the boundaries, or anywhere on the lattice, or the spontaneous infection of a site only when the absorbing state has been reached.
    Right: The infinite chain contains a continuous phase transition between an absorbing phase $\lambda > \lambda_c$ and an endemic phase $\lambda < \lambda_c$. Typical (finite size) trajectories are shown for the boundary driven system in the absorbing, critical and endemic regimes, where the horizontal direction corresponds to the length of the chain and the vertical direction is time. 
    }
    \label{fig:contactprocess}
\end{figure*}


The contact process, or SIS model, on a one-dimensional lattice is defined in terms of an array of $N$ binary variables $(n_1, \cdots,n_N)$, where $n_i \in \{0, 1\}$ indicates the occupancy of the lattice site $i$.
Occupied sites $n_i=1$ are considered infected, while empty sites $n_i = 0$ are susceptible. Infected sites can transition back to being susceptible following a Poisson process with recovery rate $\gamma$ (see Fig \ref{fig:contactprocess}). Infected sites can also infect their direct neighbors with a transmission rate $\beta$. As a result, susceptible sites become infected with a rate $k\,\beta$, where $k=0,1$ or $2$ is the number of its infected neighbors. 

After rescaling time $t \to \gamma t$, the only relevant parameter in the system is the dimensionless ratio: $\lambda = \beta/\gamma$. The probability vector over all configurations of the chain $\ket{P(t)}$ satisfies a master equation $\partial_t \ket{P(t)} = \hat W \ket{P(t)}$ with Markov generator $\hat W$, which can be written in a quantum notation \cite{SCHUTZ20011,hooyberghs2010thermodynamics,merbis2021exact} as
\begin{align}\label{generator}
	\hat W = & \, \lambda \sum_{i=1}^{N-1} \left( \hat n_i \hat w^{0\to 1}_{i+1} + \hat w^{0\to 1}_{i} \hat n_{i+1} \right)  + \sum_{i=1}^N \hat w^{1 \to 0}_{i} 
	\\ & + \hat{W}_{\rm driv}(\alpha) \nonumber
\end{align}
where $\hat n_i, \hat w_i^{0 \to 1}, \hat w_i^{1 \to 0}$ are local operators on site $i$. When represented in terms of a 2-dimensional vector space with basis states
\begin{equation}
    \ket{0} = \begin{pmatrix}
        1 \\ 0
    \end{pmatrix}\,,
    \qquad 
        \ket{1} = \begin{pmatrix}
        0 \\ 1
    \end{pmatrix}\,,
\end{equation}
these operators are $\hat n = \ket{1}\bra{1}$, $\hat w^{0 \to 1} = \ket{1}\bra{0} - \ket{0}\bra{0}$ and $\hat w^{1 \to 0} = \ket{0}\bra{1} - \ket{1}\bra{1}$, which corresponds to the matrices:
\begin{subequations}
\label{operators}
\begin{align}
\hat n = & \begin{pmatrix}
0 & 0 \\
0 & 1
\end{pmatrix}\,, \\ 
\hat w^{0 \to 1} =  & \begin{pmatrix}
-1 & 0 \\
1 & 0
\end{pmatrix} \,, \\ 
\hat w^{1\to 0} =  & \begin{pmatrix}
0 & 1 \\
0 & -1
\end{pmatrix}\,.
\end{align}
\end{subequations}
The sub-index $i$ on the operators in \eqref{generator} indicates that these operators act only on the local sites $i$. Explicit construction of the operators involves a tensor product over all sites, where only site $i$ is acted upon by the appropriate matrix in \eqref{operators}, and all other sites are multiplied by the two dimensional identity matrix. 

In the absence of the driving term $\hat{W}_{\rm driv}(\alpha)$, the Markov chain defined by $\hat{W}$ in \eqref{generator} 
is a much used example of a system with a non-equilibrium absorbing state transition (see Fig \ref{fig:contactprocess}). In the limit of infinite system size, $N \to \infty$, there is a critical value $\lambda_c$ of the control parameter $\lambda$ below which the system relaxes to the absorbing ``healthy'' state in which no site is infected. Above the critical threshold $\lambda > \lambda_c$, the system goes to a stationary (endemic) state with a non-zero density of infected sites.
The critical point of the infinitely long one-dimensional contact process is determined accurately in the literature to be $\lambda_c = 1.64896$ (see \cite{de1999precise} and references therein). As shown in Fig \ref{fig:contactprocess}, typical trajectories in the critical region display gaps in the density of all sizes.

For finite system sizes, the absorbing state is always reached, even above the critical point. It is therefore common \cite{hooyberghs2010thermodynamics} to include an additional driving process by spontaneous creation of infected sites. There are several possibilities to include this driving term. For instance, one could introduce infected sites only at the boundary of the chain with rate $\alpha$. This is done by taking the driving term $\hat{W}_{\rm driv}$ to be:
\begin{equation}\label{Wdriv}
\hat W_{\rm driv}^{\rm bdy}(\alpha) = \alpha \left(\hat w^{0\to 1}_1 + \hat{w}^{0\to 1}_N \right)\,.
\end{equation}
Alternatively, it is also possible to spontaneously occupy any empty node in the chain by including the driving term
\begin{equation}
	\hat{W}_{\rm driv}(\alpha) = \alpha \sum_{i=1}^{N} \hat{w}_i^{0\to 1}\,,
\end{equation}
where taking $\alpha \sim \frac{1}{N^2}$ ensures the effect of the driving term will vanish in the limit $N \to \infty$. This choice has the advantage of being translation invariant. A final possibility to remove the absorbing state is to occupy a random empty site only when the absorbing state has been reached, which is implemented by the driving term
\begin{equation}
	\hat{W}_{\rm driv}(\alpha) = \alpha \sum_{i=1}^n \prod_{\substack{j=1 \\ j \neq i}}^{N} \hat{v}_j \hat{w}_i^{0\to 1} \,. 
\end{equation}
Here $\hat{v} = \ket{0}\bra{0}$ is the number operator for empty sites. 

The presence of a driving term removes the absorbing state and drives the system into a Non-Equilibrium Steady State (NESS) at late times. It would therefore make more sense to talk about inactive and active regimes in the driven contact process, instead of the absorbing and endemic phases.  No known analytical solutions exist for the NESS of the driven contact process, but it is possible to obtain accurate and efficient representations of it using variational Matrix Product State methods, as we will demonstrate in this paper.

\section{MPS representation of the many-body probability vector}\label{sec:MPSrep}

The probability vector over all configuration $\ket{P(t)}$ is a $2^N$ dimensional vector, with components equal to the probability $P(\mathbf{x},t)$ of being in a certain configuration $\mathbf{x} \in \{0,1\}^N$ at time $t$. This vector can be expanded in the product basis of $N$ two-dimensional vectors $\ket{n_i}$:
\begin{equation}
\ket{P(t)} = \sum_{n_1, \ldots , n_N} P(\mathbf{x} = n_1 \ldots n_N, t) \ket{n_1 \ldots n_N}\,,
\end{equation}
with
\begin{equation}
\ket{n_1 \ldots n_N} = \ket{n_1} \otimes \ket{n_2} \otimes \ldots \otimes \ket{n_N} \,,
\end{equation}
where $\otimes$ denotes the tensor product.
We wish to approximate the high dimensional object $\ket{P(t)}$ by a lower dimensional Matrix Product State (MPS) $\ket{\Psi}$, while retaining as much information on the state of the system as possible. An MPS is a vector composed out of an array of tensors $A_i^{n_i}$
\begin{equation}\label{MPS}
	\ket{\Psi} = \sum_{n_1, \ldots , n_N=\{0,1\}} A_1^{n_1} A_2^{n_2} \ldots A_N^{n_N}  \ket{n_1  \ldots n_N} \,.
\end{equation}
Each $A_i$ is a $d\times D \times D$ dimensional rank-3 tensor, where $d$ is the physical dimension of the variables $n_i$ ($d=2$ in our case), and $D$ denotes the \emph{bond dimension} of the tensor. The tensor $A_i$ is multiplied to the left by the tensor $A_{i-1}$ along $D$ dimensions and to the right by $A_{i+1}$ along its other $D$ dimensions. The $d=2$ dimensions left corresponds to the two states of the physical variable $\ket{n_i}$. In the case of open boundary conditions, as we will use here, the first and last tensors are rank-2 (matrices) of dimension $2\times D$. The MPS hence has of the order of $N\times 2 D^2$ parameters (the total number of coefficients in the tensors), which can be much smaller than the original $2^N$ parameters of $\ket{P(t)}$, depending on the choice of $D$. 
If $D\sim 2^{N/2}$, the system is represented with all $2^N$ parameters, and no information is lost. The goal is to compress the representation of $\ket{P(t)}$ by reducing the bond dimension $D$ (and therefore the number of parameters used to represent $\ket{P(t)}$), while retaining as much information as possible about the system.

Graphically, the MPS is often depicted as a network, where the nodes represent the tensors and links are the indices. Two connected nodes are contracted over these indices. Equation \eqref{MPS} is graphically represented by:
\begin{equation}
\ket{\Psi} ={\small
\begin{array}{c}
\begin{tikzpicture}
\node[shape=circle,draw=black,fill=orange!20] (A1) at (0,0) {$A_1$};
\node[shape=circle,draw=black,fill=orange!20] (A2) at (1.5,0) {$A_2$};
\node[shape=circle] (dots) at (3,0) {$\ldots$};
\node[shape=circle,draw=black,fill=orange!20] (AN) at (4.5,0) {$A_N$};
\node[shape=circle] (g1) at (0,-1) {};
\node[shape=circle] (g2) at (1.5,-1) {};
\node[shape=circle] (gN) at (4.5,-1) {};
\path [-,line width=2pt] (A1) edge node[above] {$D$} (A2);
\path [-,line width=2pt] (A2) edge node[above] {$D$} (dots);
\path [-,line width=2pt] (AN) edge node[above] {$D$} (dots);
\path [-] (A1) edge node[left] {$d$} (g1);
\path [-] (A2) edge node[left] {$d$} (g2);
\path [-] (AN) edge node[left] {$d$} (gN);
\end{tikzpicture}
\end{array} }
\end{equation}
The open indices represent the two-dimensional physical basis of states $n_i$ at each site. The thicker internal links are the $D$ dimensional bond indices, which are contracted over in Eq.~\eqref{MPS}.

For quantum states, the bond dimension $D$ limits the amount of entanglement between sites in the chain, as the entanglement entropy $S_A(L)$ of a connected subset $A$ of $L$ sites 
is bounded by $S_A \leq 2 \log D$. Ground states of gapped, local Hamiltonians in 1D are known to obey the area law of entanglement \cite{eisert2010colloquium}, implying that $S_A$ does not scale with the volume of $A$, but only with the size of the boundary of $A$, which in 1D is a constant. For these cases the MPS provides an efficient representation of the many-body wavefunction, even for infinite systems.  In case of critical states in 1D, there is a logarithmic correction to the area law, $S_A(L) \sim \log L$ , and in this case $D$ scales polynomially with system size.  If $D$ is taken of order $2^{N/2}$, any many-body vector can be represented as an MPS.


In this work, we will use the MPS Ansatz \eqref{MPS} to find efficient approximations for many-body probability distributions over the ensemble of dynamical trajectories for the contact process. 
In that case, the bond dimension $D$ can be thought of as the effective number of relevant configurations in the model, which might be much lower than the total number of configurations in the system. Any many-body vector can be put in MPS form by a sequence of singular value decompositions (SVD) \cite{schollwock2011density} and 
$D$ represents the number of singular values kept between the bonds after truncating the smaller singular values. The SVD constitutes a linear transformation on the basis states (in this case: the configurations of the chain) and by truncating the bonds to the largest singular values, only the most relevant linear combinations of basis states are kept.

\section{Thermodynamics of dynamical trajectories and large deviation statistics}\label{sec:thermo}

The time-evolving stochastic process traces out a \emph{trajectory}  $\omega_{t_k} = \{\mathbf{x}_0(t_0), \mathbf{x}_1(t_1), \ldots , \mathbf{x}_k(t_k) \}$, by hopping to configuration $\mathbf{x}_l$ at hopping time $t_l$. 
To study the statistics of the trajectories in the system, we consider the dynamical activity $K(\omega_t)$, defined as the total number of configuration changes along a trajectory $\omega_t$ of duration~$t$ \cite{lecomte2007thermodynamic,garrahan2007dynamical}. The dynamical activity is an order parameter for the phase transition between active and inactive trajectories; this transition is dynamical in nature. 
Below, we will review how the moment generating function for the dynamical activity, in the ensemble of all possible trajectories, can be expressed as the norm of a tilted probability vector. 

\subsection{Dynamical activity}

The dynamical activity $K(\omega_t)$ is a trajectory dependent random variable, which forces us to track the probability vector $\ket{P(k,t)}=\sum_{\mathbf{x}} P(\mathbf{x}, k,t) \ket{\mathbf{x}}$ that the system is in any configuration $\mathbf{x}$ and has activity $k$ at time $t$. Thus $P(\mathbf{x}, k,t) = \bracket{\mathbf{x}}{P(k,t)}$. 
The moment generating function $\Z_K(s)$ for the dynamical activity is
\begin{equation}
	\Z_K(s) = \langle e^{s K(\omega_t)} \rangle_{\omega_t}  = \sum_{k=0}^\infty \langle \mathbf{1} | e^{s \,k} \ket{P(k,t)}\,.
\end{equation}
Here the brackets $\langle \, \rangle_{\omega_t}$ denote the ensemble average over all dynamical trajectories $\omega_t$ leading up to time $t$ from an initial configuration $\mathbf{x}_0 \in \{0,1\}^N$. The unit column vector $\bra{ \mathbf{1}} $ is the \emph{flat state}, such that the contraction of $\bra{\mathbf{1}}$ with a many-body vector returns the $L^1$--norm of that vector.

The moment generating function $\Z_K(s)$ is more conveniently computed as the $L^1$-norm of a \emph{tilted probability vector} $\ket{P(s,t)}$, defined as the Laplace transform of $\ket{P(k,t)}$ \cite{garrahan2009first}:
\begin{subequations}
\begin{align}
	\Z_K(s)  & = \bracket{\mathbf{1}}{P(s,t)}\,,\\
	\ket{P(s,t)} & = \sum_k e^{s \,k} \ket{P(k,t)} \,.
\end{align}
\end{subequations}
The tilted probability vector evolves in time by the action of a tilted generator $\tilde{W}(s)$, which can be obtained from the Markov generator \eqref{generator} by multiplying all off-diagonal elements with $e^s$ \cite{lecomte2007thermodynamic,garrahan2007dynamical,garrahan2009first,gorissen2009density,hooyberghs2010thermodynamics},
\begin{equation}
	\partial_t \ket{P(s,t)} = \tilde{W}(s) \ket{P(s,t)}\,,
\end{equation}
with
\begin{align}\label{tiltedgenerator}
\tilde W(s) = & \lambda \sum_{i=1}^{N-1} \left( \hat n_i \tilde w^{0\to 1}_{i+1}(s) + \tilde w^{0\to 1}_{i}(s) \hat n_{i+1} \right) \nonumber \\
& + \sum_{i=1}^N \tilde w^{1 \to 0}_{i}(s) + \tilde{W}_{\rm driv}(\alpha, s)\,,
\end{align}
such that
\begin{equation}
\tilde w^{0 \to 1}(s) = \begin{pmatrix}
-1 & 0 \\
e^s & 0
\end{pmatrix} \,, \quad 
\tilde w^{1\to 0}(s) = \begin{pmatrix}
0 & e^s \\
0 & -1
\end{pmatrix}\,.
\end{equation}
At $s=0$, the tilted generator reduces to the Markov generator \eqref{generator}. Away from $s=0$, the tilted generator is neither Hermitian nor stochastic, so it will not conserve total probability. For stochastic processes with local detailed balance, there does exist a similarity transformation to a Hermitian matrix \cite{garrahan2009first,banuls2019using}. In case of the contact process, the infection transmission requires nearest neighbor interactions, while recovery is a purely local process, and hence the process does not satisfy local detailed balance. 

The exponential tilting weighs trajectories by their activity. When $s>0$, active trajectories are exponentially favored, while for $s<0$ they are exponentially suppressed. In the case of the boundary driven system, the tilted generator becomes diagonal in the extreme inactive limit $s \to -\infty$, with leading eigenvalue $-2\alpha$, due to the boundary driving term. The corresponding eigenvector is the completely healthy state: the product state with all sites in the $\ket{0}$ state. $\ket{P(s,t)}$ reduces to the time-dependent probability vector $\ket{P(t)}$ at $s=0$, which at late times becomes the NESS of the driven system. This is not a product state, but contains correlations between the sites in the chain. No known analytical solution for the NESS exists in the literature to the best of our knowledge. 

\subsection{Large deviation principle}
For the driven contact process, the absorbing state is eliminated and the system is driven into a NESS. In that case, the probability distribution $P(k,t)$ of observing activity $k$ at time~$t$ satisfies a Large Deviation (LD) principle, such that at late times: $P(k,t) \sim e^{-t \phi(k/t)}$, with $\phi(k)$ the LD rate function \cite{touchette2009large}. The moment generating function $\Z_K(s)$ similarly satisfies a LD principle:
\begin{equation}
	\Z_K(s) \sim e^{t \Theta(s)} \,, \quad \Theta(s) = \lim_{t\to \infty} \frac{1}{t} \log \langle e^{s\, K(\omega_t)} \rangle \,.
\end{equation}
The \textit{Scaled Cumulant Generating Function} (SCGF) $\Theta(s)$ plays the role of dynamical free energy 
and is related to the LD rate function $\phi(k)$ by Legrendre transform \cite{touchette2009large}
\begin{equation}\label{ratefunction}
	\phi(k) = \sup_s \left( s\,k - \Theta(s) \right)\,.
\end{equation}
As the tilted probability vector is $\ket{P(s,t)} = e^{\tilde W(s)t} \ket{x_0}$, the SCGF becomes the leading eigenvalue $\lambda_0$ of the tilted generator $\tilde{W}(s)$,
\begin{equation}
	\Theta(s) = \lambda_0 \left(\tilde{W}(s) \right)\,.
\end{equation}
Here, the similarity between the tilted generator and a many-body Hamiltonian comes into play. Using the MPS Ansatz for the many-body probability vector $\ket{P}$, we can obtain the leading eigenvalue of $\tilde{W}(s)$ by a variational search over MPS states:
\begin{equation}\label{dfe}
	\Theta(s) = \max_{\ket{\Psi}} \frac{\bra{\Psi} \tilde{W}(s) \ket{\Psi}}{\bracket{\Psi}{\Psi}}
\end{equation} 
where $\ket{\Psi}$ is a many-body vector and $\bra{\Psi}$ its complex conjugate. In the coming sections we will use variational MPS algorithms to find efficient representations for the leading eigenvector of $\tilde{W}(s)$ and use this to compute the NESS, the SCGF and other observables directly in the ensemble average over all trajectories.

\section{NESS computation by variational MPS}
\label{sec:varNESS}

Here we show how the NESS of the 1D contact process can be found by a variational search over MPS states \eqref{MPS}.
We comment on the difference with sampling based (Monte-Carlo) methods and demonstrate the accuracy of our approximations. 
To illustrate the effectiveness of the MPS representation, we compute the distribution of gaps of length $k$ in the NESS, including the expectation values for large gaps, which are difficult to obtain accurately using sampling-based methods.

\subsection{Variational MPS methods}
The NESS is the leading right eigenvector of the Markov generator \eqref{generator}, which corresponds to an eigenvalue of $0$. The eigenvector is found in MPS form by sequentially optimizing for a single tensor $A_i$ in \eqref{MPS}, while keeping all other tensors constant, and iteratively sweeping over the chain until convergence. There are many good reviews detailing variational MPS algorithms for many-body quantum systems, such as \cite{verstraete2008matrix,schollwock2011density,orus2014practical}. Here, we will focus mainly on the differences with the usual MPS algorithms arising from the fact that we are optimizing for a probability distribution and not a complex-valued wavefunction.

The Markov generator $\hat{W}$ is not a Hermitian matrix, and hence its left and right eigenvectors are not related to each other by complex conjugation. 
This does, however, not pose any serious problem, as long as we consistently update the MPS using the right eigenvectors of $\hat{W}$. The Markov generator is first represented as a Matrix Product Operator (MPO), with constant bond dimension $D_W =4$ (see \cite{schollwock2011density} for a review on the construction of MPOs). Schematically, this is done by representing $\hat{W}$ as a chain of rank-4 tensor of dimensionality $(2,2,4,4)$, except at the boundaries of the chain, where the tensors are of rank 3. In the case of the boundary driven system, the MPO is constructed as follows:

\begin{align}
	\hat{W}  & =  L M \dots M R   \\ \nonumber
&	= {\small
		\begin{array}{c}
			\begin{tikzpicture}
				\node[shape=rectangle,draw=black,fill=green!20] (A1) at (0,0) {$L$};
				\node[shape=rectangle,draw=black,fill=green!20] (A2) at (1,0) {$M$};
				\node[shape=circle] (dots) at (2.25,0) {$\ldots$};
				\node[shape=rectangle,draw=black,fill=green!20] (ANm1) at (3.5,0) {$M$};
				\node[shape=rectangle,draw=black,fill=green!20] (AN) at (4.5,0) {$R$};
				\node[shape=circle] (g1) at (0,-1) {};
				\node[shape=circle] (g2) at (1,-1) {};
				\node[shape=circle] (gNm1) at (3.5,-1) {};
				\node[shape=circle] (gN) at (4.5,-1) {};
				\node[shape=circle] (g1up) at (0,1) {};
				\node[shape=circle] (g2up) at (1,1) {};
				\node[shape=circle] (gNm1up) at (3.5,1) {};
				\node[shape=circle] (gNup) at (4.5,1) {};
				\path [-,line width=2pt] (A1) edge node[above] {$4$} (A2);
				\path [-,line width=2pt] (A2) edge node[above] {$4$} (dots);
				\path [-,line width=2pt] (ANm1) edge node[above] {$4$} (dots);
				\path [-,line width=2pt] (ANm1) edge node[above] {$4$} (AN);
				\path [-] (A1) edge node[left] {$2$} (g1);
				\path [-] (A2) edge node[left] {$2$} (g2);
				\path [-] (ANm1) edge node[left] {$2$} (gNm1);
				\path [-] (AN) edge node[left] {$2$} (gN);
				\path [-] (A1) edge node[left] {$2$} (g1up);
				\path [-] (A2) edge node[left] {$2$} (g2up);
				\path [-] (ANm1) edge node[left] {$2$} (gNm1up);
				\path [-] (AN) edge node[left] {$2$} (gNup);
			\end{tikzpicture}
	\end{array} }
\end{align}
with:
\begin{equation}
M = \begin{pmatrix}
	\mathbb{1} & 0 & 0 & 0 \\
	 \lambda \hat{w}^{0\to 1} & 0 & 0 & 0 \\
	  \lambda \hat{n}  & 0 & 0 & 0 \\
	\hat{w}^{1\to 0}  & \hat{n} & \hat{w}^{0\to 1} & \mathbb{1}
\end{pmatrix}\,,
\end{equation}
and 
\begin{align}
		L = & \begin{pmatrix}
		\hat{w}^{1\to 0} + \alpha \hat{w}^{0 \to 1} & \hat{n} & \hat{w}^{0\to 1} & \mathbb{1}
	\end{pmatrix} \,, \\
R = & \begin{pmatrix}
	\mathbb{1} \\  \lambda \hat{w}^{0\to 1} \\ \lambda \hat{n}  \\ \hat{w}^{1\to 0} + \alpha \hat{w}^{0 \to 1} 
\end{pmatrix} \,.
\end{align}

The single tensor optimization problem for site $i$ is then constructed by removing the local tensors $A_i$ and $A^\dagger_i$ from the contraction $\bra{\Psi} \hat{W} \ket{\Psi}$, which creates a $D^2d$ dimensional matrix $\hat{W}_{\rm eff}^i$. For instance, for $N=5$ and $i=3$, the effective generator $\hat{W}_{\rm eff}$ is:
\begin{equation}\label{Weffcontraction}
	\hat{W}_{\rm eff} = 		
		\begin{array}{c}
		\begin{tikzpicture}
			\node[shape=rectangle,draw=black,fill=green!20] (L) at (0,0) {$L$};
			\node[shape=rectangle,draw=black,fill=green!20] (M2) at (1,0) {$M$};
			\node[shape=rectangle,draw=black,fill=green!20] (M3) at (2,0) {$M$};
			\node[shape=rectangle,draw=black,fill=green!20] (M4) at (3,0) {$M$};
			\node[shape=rectangle,draw=black,fill=green!20] (R) at (4,0) {$R$};			
			\node[shape=circle,draw=black,fill=orange!20] (A1) at (0,1) {{\scriptsize $A_1$}};
			\node[shape=circle,draw=black,fill=orange!20] (A2) at (1,1) {{\scriptsize $A_2$}};
			\node[shape=circle] (A3) at (2,1) {};
			\node[shape=circle,draw=black,fill=orange!20] (A4) at (3,1) {{\scriptsize $A_4$}};
			\node[shape=circle,draw=black,fill=orange!20] (A5) at (4,1) {{\scriptsize $A_5$}};
					`
			\node[shape=circle,draw=black,fill=orange!20] (Ad1) at (0,-1) {{\scriptsize $A_1$}};
			\node[shape=circle,draw=black,fill=orange!20] (Ad2) at (1,-1) {{\scriptsize $A_2$}};
			\node[shape=circle] (Ad3) at (2,-1) {};
			\node[shape=circle,draw=black,fill=orange!20] (Ad4) at (3,-1) {{\scriptsize $A_4$}};
			\node[shape=circle,draw=black,fill=orange!20] (Ad5) at (4,-1) {{\scriptsize $A_5$}};
			\path [-,line width=2pt] (L) edge node[above] {} (M2);
			\path [-,line width=2pt] (M2) edge node[above] {} (M3);
			\path [-,line width=2pt] (M3) edge node[above] {} (M4);
			\path [-,line width=2pt] (M4) edge node[above] {} (R);
			\path [-] (L) edge node[left] {} (Ad1);
			\path [-] (M2) edge node[left] {} (Ad2);
			\path [-] (M3) edge node[left] {} (Ad3);
			\path [-] (M4) edge node[left] {} (Ad4);
			\path [-] (R) edge node[left] {} (Ad5);
			\path [-] (L) edge node[left] {} (A1);
			\path [-] (M2) edge node[left] {} (A2);
			\path [-] (M3) edge node[left] {} (A3);
			\path [-] (M4) edge node[left] {} (A4);
			\path [-] (R) edge node[left] {} (A5);
			\path [-,line width=2pt] (A1) edge node[above] {} (A2);
			\path [-,line width=2pt] (A2) edge node[above] {} (A3);
			\path [-,line width=2pt] (A3) edge node[above] {} (A4);
			\path [-,line width=2pt] (A4) edge node[above] {} (A5);
			\path [-,line width=2pt] (Ad1) edge node[above] {} (Ad2);
			\path [-,line width=2pt] (Ad2) edge node[above] {} (Ad3);
			\path [-,line width=2pt] (Ad3) edge node[above] {} (Ad4);
			\path [-,line width=2pt] (Ad4) edge node[above] {} (Ad5);
		\end{tikzpicture}
	\end{array}
\end{equation}

The tensor $A_i$ in the MPS is then replaced with the leading right eigenvector of $\hat{W}_{\rm eff}^i$. This vector is found by the implicitly restarted Arnoldi method \cite{lehoucq1998arpack}, iteratively optimizing for the eigenvector corresponding to the largest real eigenvalue. In practice, it is computationally more efficient to contract $\hat{W}_{\rm eff}^i$ with the local tensor $A_i$ in the iterative eigenvector search directly, instead of explicitly constructing the $dD^2$ dimensional matrix \eqref{Weffcontraction}. In this way, the eigenvector search comes at a computational cost which scales as $D^3$. It is also beneficial to form the optimization problem for two sites simultaneously. The resulting effective generator is then a $d^2D^2$ dimensional matrix and its eigenvector can be brought back to the MPS form by performing a singular value decomposition (SVD). After the SVD, we truncate the singular value spectrum by dropping all singular values below a fixed threshold $\sim 10^{-16}$. This procedure dynamically adjusts the bond dimension and keeps only singular values which are relevant. In practice, we need to provide also an upper bound on the bond dimension, usually taken to be around $D_{\rm max} \sim 200$, to prevent the bond dimension to become too large. There are hence two ways to dynamically control the bond dimension of the MPS, one by truncating the SV spectrum on each bond, another by controlling the maximal bond dimension for each bond.

\begin{figure*}[ht]
	\includegraphics[width=0.9\linewidth]{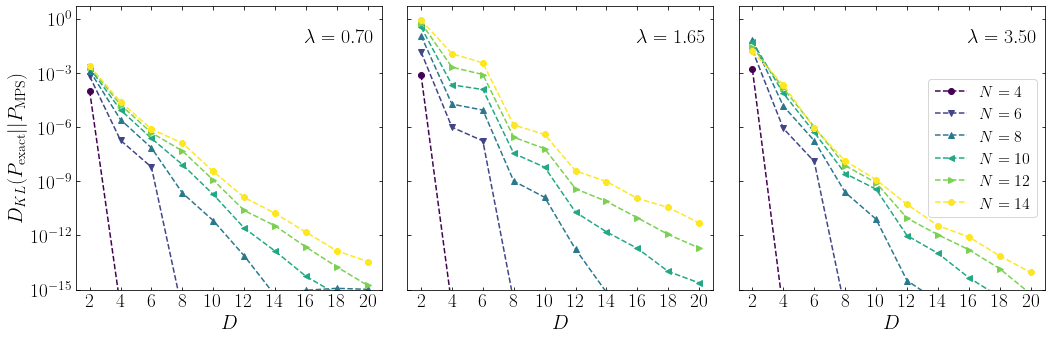}
	\caption{\label{fig:KLdiv} 
 The Kullback-Leibler divergence between the NESS found as leading eigenvector of the explicit construction of the $2^N$ dimensional Markov generator $P_{\rm exact}$ and the MPS representation of the NESS $P_{\rm MPS}$ for various values of $D$ and $N$ and $\lambda = 1$ (left), $\lambda \simeq \lambda_c $ (middle) and $\lambda = 3$ (right). Parameters for this plot are $\epsilon = 10^{-8}$ and $\alpha = \frac{1}{N^2}$ with spontaneous infection at all empty sites.}
\end{figure*}

\begin{figure*}[ht]
	\includegraphics[width=0.9\linewidth]{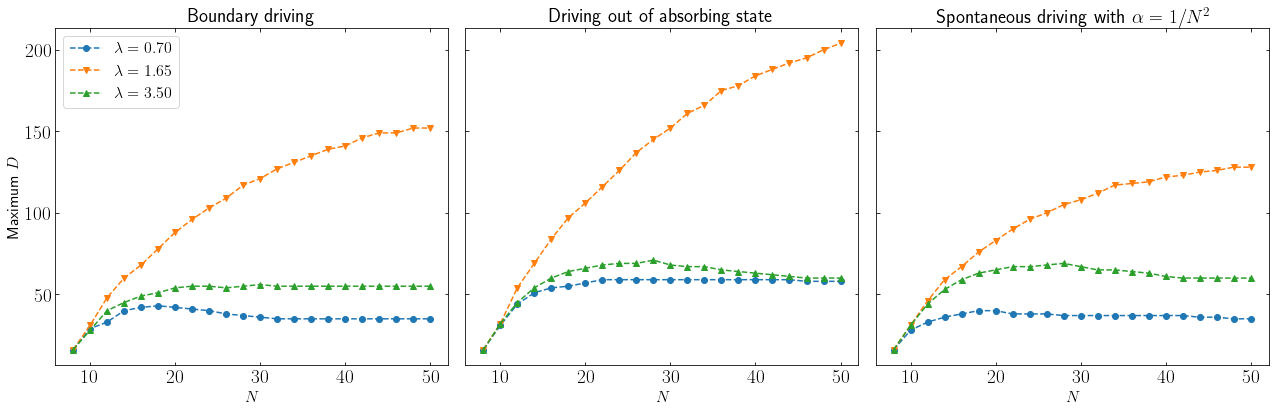}
	\caption{\label{fig:DvsN} The dynamically obtained maximal bond dimension $D$ of the MPS representation of the NESS, obtained by truncating singular values below the threshold value of  $10^{-14}$, as a function of system size $N$ for three values of $\lambda$ and three different driving protocols. The three values for $\lambda$ were chosen to lie in the absorbing phase ($\lambda= 0.7$), the active phase ($\lambda = 3.5$) and in the critical region $\lambda = \lambda_c = 1.649$. The critical value of $\lambda = \lambda_c$ is the critical value for the system in the thermodynamic limit without driving. The center figure is the closest to that system and displays the fastest increase in bond dimension close to criticality. The convergence criteria on the MPS was set to $\epsilon = 10^{-8}$ for these plots.}
\end{figure*}

One iterative sweep across the MPS starts with sequentially replacing pairs of local tensors $A_i$, $A_{i+1}$, starting from site $(1,2)$ to the end, followed by sequential pairwise updates in the other direction. The backward sweep is done for complementary pair combinations with regard to the forward sweep, i.e. if in the forward sweep sites $(1,2), (3,4) , \ldots $ are paired together, the backward sweep pairs the sites $\ldots, (5,4), (3,2)$ followed with a single site update of the first site. By repeatedly sweeping across the chain in this fashion, the MPS Ansatz converges to the leading right eigenvector $\ket{\Psi_0}$, which is the NESS of the Markovian process. 

Once the MPS representation $\ket{\Psi_0}$ is found, we can use it to compute observables. It is important to note that $\ket{\Psi_0}$ is directly corresponding to a probability distribution (and not its $L^2$ norm), but the optimization procedure does not produce a normalized vector. Therefore, the probability distribution $P(\mathbf{x})$ over all configurations $\mathbf{x}$ in the NESS, is given as:
\begin{equation}
    P(\mathbf{x}) = \frac{1}{\bracket{\mathbf{1}}{\Psi_0} } \ket{\Psi_0}\,.
\end{equation}
Likewise, we compute observables in the $L^1$ norm, while dividing by the $L^1$ norm of the total vector $\bracket{\mathbf{1}}{\Psi_0}$. 
For instance, the expected density in the chain is computed as:
\begin{equation}\label{densexp}
	\vev{n} = \frac{\bra{\mathbf{1}} \frac{1}{N} \sum_{i=1}^{N} \hat{n}_i \ket{\Psi_0} }{ \bracket{\mathbf{1}}{\Psi_0}} \,,
\end{equation}
and similarly for other observables. The contraction is implemented efficiently by representing the observable of interest as an MPO, and performing a horizontal contraction of the MPO with the state $\ket{\Psi_0}$ on top and a direct product of local flat states $\bra{\mathbf{1}} = (1,1)$ on the bottom. If one first contracts the lower physical index of the MPO with the flat state, followed by the contraction of the resulting tensor with the local tensors $A_i$, this can be done with a computational cost scaling as $\mathcal{O}(ND_{MPO}D^2)$, where $D_{MPO}$ is the bond dimension of the MPO.

The local optimization process is repeated until the density in the chain has converged, meaning that the relative difference between $\vev{n}_k$ after the $k$-th sweep and $\vev{n}_{k-1}$ has dropped below a threshold value: $\left| \frac{\vev{n}_{k-1} - \vev{n}_{k}}{\vev{n}_{k}}\right| < \epsilon$, with $\epsilon = 10^{-6} - 10^{-8}$ depending on system size. 
To demonstrate the accuracy of the MPS representation as a function of the bond dimension $D$, we show in Fig.~\ref{fig:KLdiv} the Kullback-Leibler (KL) divergence between $P_{\rm exact}$, obtained as the leading right eigenvector of the explicit construction of the $2^N$ dimensional Markov generator, and $P_{\rm MPS}$ obtained from the contraction of the MPS representation, for small system sizes and various values of the maximal bond dimension $D$ and transmission rate $\lambda$. With bond dimensions $D = 2^{N/2}$, the MPS is exact and the KL divergence with the exact solution becomes $O(10^{-16})$, within machine precision. For smaller bond dimension the MPS representation is still very close to the exact solution, however, larger system sizes do require a larger bond dimension to achieve the same accuracy, especially near the critical transmission rate $\lambda_c$. 

In Fig.~\ref{fig:DvsN} we investigate the growth of the bond dimension with system size $N$ for the three different driving protocols and three values of the transmission rate $\lambda$. We let the algorithm dynamically adjust the bond dimensions for each bond, by truncating the singular values below a lower threshold, taken to be $10^{-14}$.  The figure displays the maximal bond dimension in the chain as a function of system size and we observe that in the active and inactive region, the bond dimension tends to a constant value for most driving protocols. At the critical point, the bond dimension is expected to grow linearly in $N$, due to the emergent scale invariance \cite{calabrese2004entanglement}. In Fig \ref{fig:DvsN} the orange curves are close to criticality, but not exactly at the critical point due to finite size and driving effects, hence a scaling slower than linear is observed here. In any case, a growth of $D$ which is at most linear in $N$ ensures the efficiency of the MPS representation of the NESS, compared to the maximal exponential growth of $D$ with system size. By comparing with Fig. \ref{fig:KLdiv}, we see that the KL divergence with the exact solution is extremely small ($< 10^{-12}$) for the smaller system sizes ($N \simeq 8-14$). This shows that the MPS representation is also accurate. 




\subsection{Distribution of gaps in the NESS}


\begin{figure*}[t]
	\includegraphics[width=0.48\linewidth]{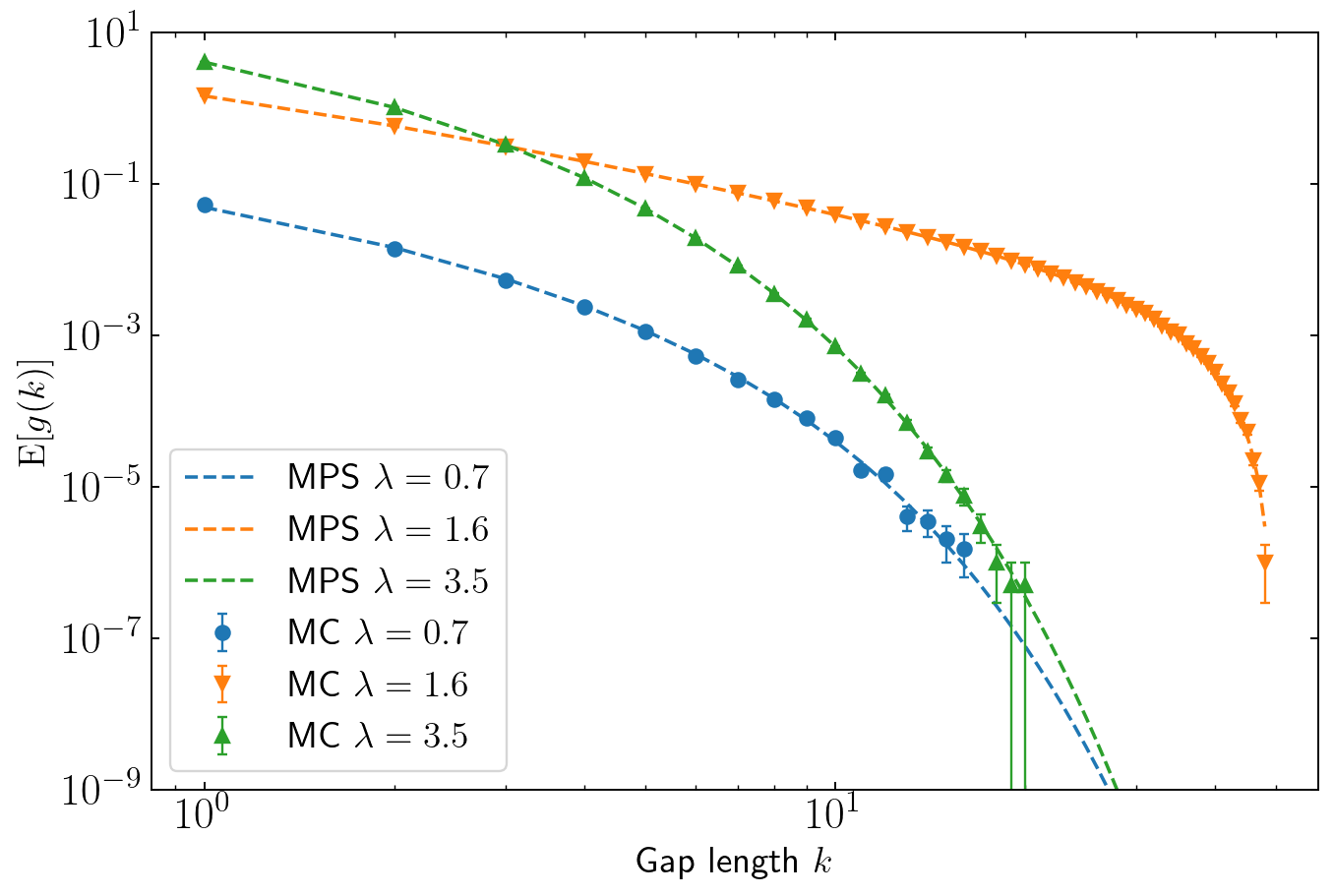}
	\includegraphics[width=0.48\linewidth]{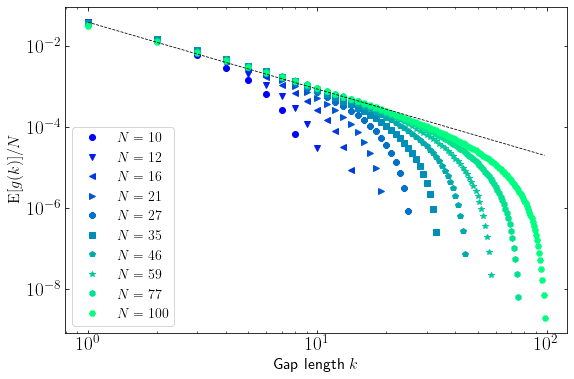}
	\caption{\label{fig:Gaps} Left: The distribution of gaps of length $k$ in a chain of length $N=50$ at three different values of $\lambda$. The dotted lines are obtained from first finding an MPS representation of the NESS, followed by the contraction of this MPS with an MPO which gives the expected number of gaps of length $k$. Solid points with error bars are obtained from dynamical Monte-Carlo sampling of over $2 \times 10^6$ trajectories of the physical process, recorded once the system as reached stationarity. Right: The distribution of gaps at $\lambda_c$ for various system sizes. The dotted line is a power law fit for small $k$. }
\end{figure*}

By searching for an efficient MPS representation, we are approximating directly the full $2^N$ dimensional probability vector in the NESS. Once the optimization process has finished to the desired accuracy, we can, in principle, compute efficiently any observable we want directly in the ensemble average over all configurations. This includes global and local densities, variances thereof, but also all of the higher moments, the complete moment generating function for the observable, or marginals, by projecting the probability vector on specific microscopic configurations.

This is a fundamentally different approach to more common sampling-based methods, frequently employed for the analysis of such systems. Using these methods, one typically needs to generate a large number of trajectories and then compute the observables of interest as the empirical averages over the set of generated trajectories. This makes the accurate computation of expectation values of rare events very demanding, as it requires one to generate a large set of trajectories to even encounter the rare event once. Using the MPS method, computing expectation values for events which occur very infrequently is not more difficult than any other event, as we have access to an approximation of the full probability distribution over all possible configurations of the system.

To illustrate this point with an example pertaining to the current dynamical system, we use our MPS representation of the NESS to compute the distribution of gaps of length $k$ in a chain of total length $N$.  A gap of length $k$ is defined as a configuration with $k$ empty sites in a row, preceded and followed by an occupied site. The expected number of gaps of length $k$ in a chain of length $N$ can be computed from the MPS representation of the NESS by contracting it with an MPO of bond dimension $k+3$, constructed as follows:
\begin{equation}\label{gapMPO}
	\hat{G}_k =  L_k M_k \ldots M_k R_k 
\end{equation}
with:
\begin{equation}
	M_k = \begin{pmatrix}
		\mathbb{1} & \hat{n} & 0 & \ldots &  &0  \\
		0 & 0 & \hat{v} &  &  \\
		 &  &  & \ddots & & \vdots \\
		\vdots &  &  & &  \hat{v} & 0 \\
		 &  &  & & & \hat{n} \\
		 0 &  &  & \hdots & 0 & \mathbb{1}
	\end{pmatrix}\,,
\end{equation}
and 
\begin{equation}
	L_k = \begin{pmatrix}
		\mathbb{1} & \hat{n} & 0 & \ldots & 0
	\end{pmatrix} \,, \qquad
	R_k = \begin{pmatrix}
		0 \\  \vdots \\ 0  \\ \hat{n} \\ \mathbb{1}
	\end{pmatrix} \,.
\end{equation}
When explicitly contracting the MPO \eqref{gapMPO}, one obtains:
\begin{equation}
	\hat{G}_k = \sum_{i=1}^{N-1-k} \hat{n}_i \hat{v}_{i+1} \ldots \hat{v}_{i+k+1} \hat{n}_{i+k+2} \,.
\end{equation}
This operator exactly projects the state on all possible positions where a gap of length $k$ could be located in the chain.

In Fig.~\ref{fig:Gaps} we show the distribution of gaps in a chain of length $N=50$ computed by first optimizing an MPS for the NESS at a given $\lambda$, and then contracting the MPO \eqref{gapMPO} with the MPS. For this figure we have used a driving protocol only on the left boundary of the chain. We notice that for sub- and super-critical values of $\lambda$, the expectation values for gaps drops exponentially with the length of the gap. For $\lambda \sim \lambda_c$ the expected number of gaps remains larger for large $k$ and follows a power law initially. For larger gaps, the expected value drops again exponentially, which is due to finite size effects, as the gap size starts to become close to the length of the chain. The right panel of Fig.~\ref{fig:Gaps} shows that the expected number of gaps over $N$ at $\lambda = \lambda_c$ fits a power law  $\mathrm{E}[g(k)] \sim k^{-\nu}$, which is characteristic for scale-invariance at criticality. We find an exponent of $\nu = 1.71(6)$, which corresponds to the fractal dimension \cite{hovi1996gap, ginelli2005directed} for the critical contact process.

In the left panel of Fig.~\ref{fig:Gaps}, we compare the expected number of gaps with ensemble averages over trajectories generated by Markov Chain Monte Carlo methods. The solid points with error bars are data obtained by averaging the observed number of gaps in $2 \times 10^6$ configurations generated by Gillespie's algorithm \cite{gillespie1977exact}. The sampling-based algorithm gives excellent agreement with the MPS results for small gap sizes. Large gaps (of the order of 10-20 sites) are, however, extremely rare in both the absorbing and the active phase, making it difficult to obtain accurate statistics using sampling-based methods. The MPS, however, has no problem in providing an accurate prediction in the expected number of gaps in these cases. 
We find these results are robust under varying the maximal bond dimension $D$, such that finite $D$ effects are irrelevant for this observable. 


\section{LD statistics from variational MPS}
\label{sec:varMPS}

In this section we will compute the scaled cumulant generating function (SCGF) (or, the dynamical free energy) $\Theta(s)$ by a variational search over MPS states \eqref{MPS}, while dynamically adjusting the bond dimension for each bond in the chain. We will use this to study the SCGF and other observables of interest, such as the dynamical activity at finite $s$ and the susceptibility $\chi(s) = \partial_s^2 \Theta(s)$, whose peak characterizes the active/inactive transition for finite system sizes. The results show a limiting behavior consistent with a discontinuous phase transition in dynamical activity between the active and inactive phases, as a function of $s$, for any value of $\lambda$. The dynamical system hence contains both a continuous phase transition in the density of the stationary state (while varying $\lambda$) and a discontinuous phase transition in the dynamical activity of trajectories (while varying $s$). 

\subsection{MPS methods for the SCGF}

In order to obtain an MPS representation which is both accurate and efficient, we have to find the optimal bond dimension $D$. This should be just big enough to give accurate predictions, but as small as possible to optimize computational efficiency. The goal in this section is to obtain the leading eigenvector $\Theta(s)$ of the tilted generator from \eqref{dfe}. We do this by dynamically adjusting the bond dimension for each bond, as described in the previous section. In this case, however, the convergence criteria on the MPS is based on the variance in $\Theta(s)$, given by ${\rm Var}[\Theta(s)] = \bra{\mathbf{1}} \tilde{W}(s)^2 \ket{\Psi(s)} - \Theta(s)^2$. When this drops below a threshold value, taken to be $10^{-8}$, we can assure the accuracy of the MPS. After having found the leading right eigenvector of $\tilde{W}(s)$ in MPS form for a specific value of $s$, this state is used as the initial state for the next value of $s$. 


\begin{figure}[t]
	\includegraphics[width=\columnwidth]{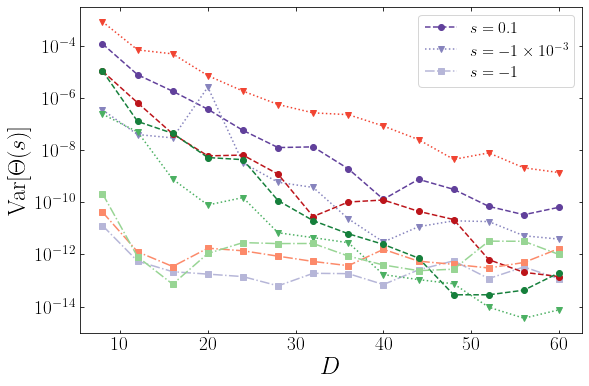}
	\caption{\label{fig:VarvsD} The variance in $\Theta(s)$ as a function of the bond dimension $D$ for various values of $s$. Different colors indicate different $\lambda$ values: Purple lines correspond to $\lambda =1$, reds to $\lambda = \lambda_c$ and greens to $\lambda = 3.5$. This plot was obtained for $N=50$, using boundary driving.}
\end{figure}


We investigate the variance as a function of the maximal bond dimension $D_{\rm max}$ for $N=50$ and various values of $\lambda$ and $s$ in Fig.~\ref{fig:VarvsD}. Here, the SVD cut-off was removed, such that all bonds in the MPS have bond dimension $D_{\rm max}$. This shows that even small bond dimensions give a good approximation in most cases, particularly when $\lambda$ and $s$ are chosen such that the process is in the inactive phase.

\begin{figure}[!th]
	\includegraphics[width=\columnwidth]{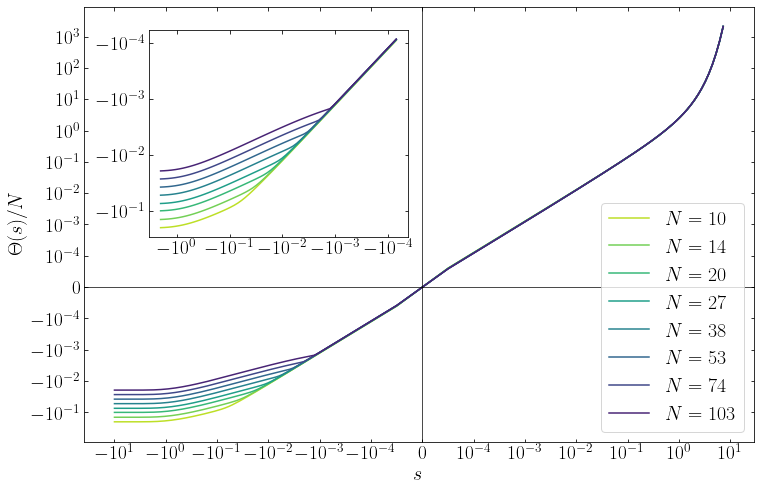}
	\includegraphics[width=\columnwidth]{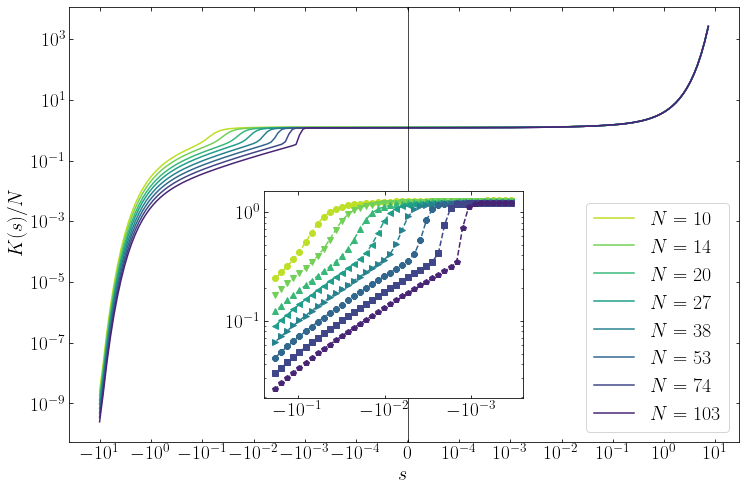}
	\includegraphics[width=\columnwidth]{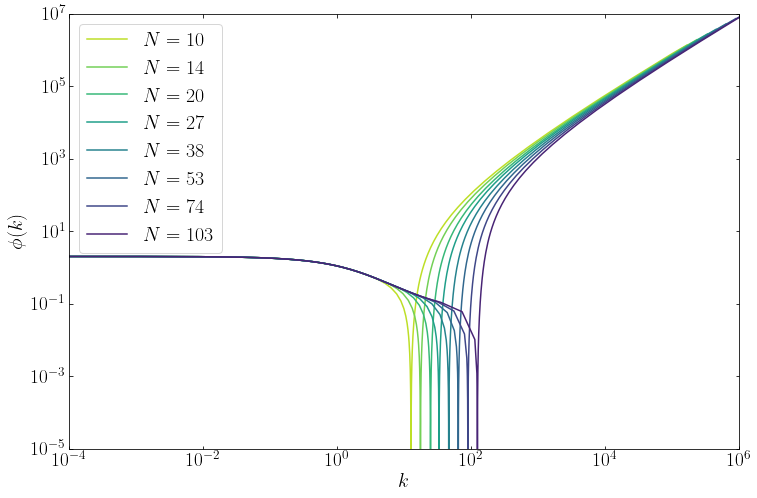}
	\caption{\label{fig:FEresults} The scaled cumulant generating function (top) normalized by system size for various $N$ at $\lambda = 2 > \lambda_c$ using the boundary driving protocol. The middle panel shows the activity per site as a function of $s$, obtained as $K(s)/N = \partial_s \Theta(s)/N$. The rate function (bottom) clearly shows zeros at the expected value for the activity. Due to the boundary driving, inactive trajectories (small $k$) all have the same constant decay rate $\phi(k)$.}
\end{figure}

\begin{figure*}[t]
	\includegraphics[height=0.24\textwidth]{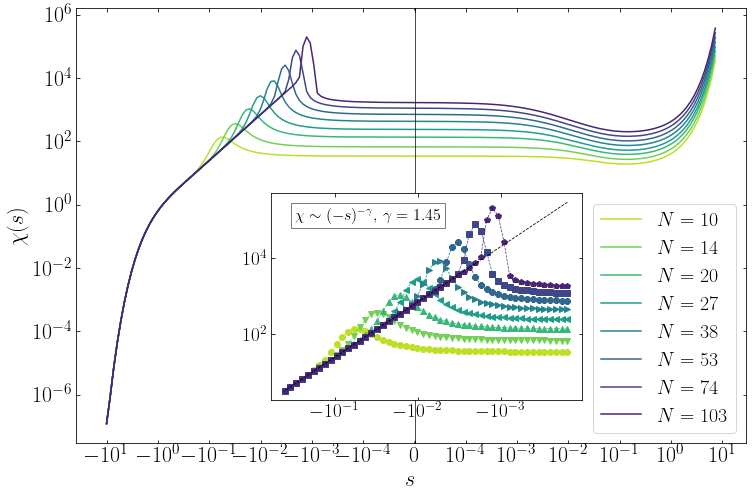}
	\includegraphics[height=0.24\textwidth]{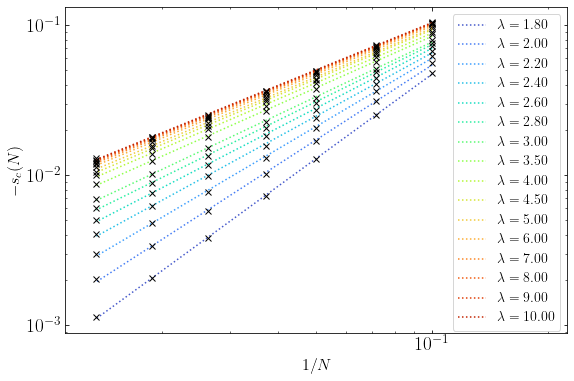}
	\includegraphics[height=0.24\textwidth]{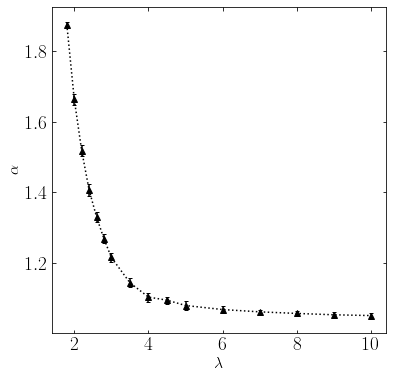}
	\caption{\label{fig:criticalsN} Left: The susceptibility $\chi(s) = \partial_s^2 \Theta(s)$ shows a peak for finite $N$ at a critical value of $s = s_c(N)$ which becomes more narrow for increasing $N$. Middle: The critical values $s_c(N)$ at different values for the infection rate $\lambda > \lambda_c$. The results show a scaling of $s_c(N) \sim - N^{-\alpha} $ with $\alpha > 1$. Right: The exponents $\alpha$ as a function of infection rate $\lambda$.}
\end{figure*}

\subsection{Critical threshold and finite size scaling}

In Fig.~\ref{fig:FEresults} we show $\Theta(s)/N$ for various system sizes at transmission rate $\lambda = 2 > \lambda_c$. We see that for trajectories tilted towards the active region ($s>0$), all lines lie perfectly on top of each other. The curves all pass through the origin, as at this point the tilted generator becomes the Markov generator with leading eigenvalue equal to zero. Around $s = 0$, the SCGF grows linearly in both $s$ and $N$, i.e. $\Theta(s) \sim  s N$. For large negative $s$, the SCGF converges to $-2$, as for inactive trajectories the only activity comes from the boundary driving terms. At a critical value for $s = s_c(N)$, a kink in the free energy appears, leading to a jump in the activity $K(s) = \partial_s\Theta(s)$, illustrated in the middle panel of Fig. \ref{fig:FEresults}. This is characteristic of a discontinuous phase transition. For $\lambda > \lambda_c$, the critical value $s_c(N)$ converges to zero in the large $N$ limit. The activity $K(s)$ gives the expectation value of the dynamical activity $K$ at $s=0$, and away from $s=0$ it represents the activity in the tilted trajectories. The lower panel shows the rate function $\phi(k)$, computed by performing the Legendre transform \eqref{ratefunction}. Its value gives the rate of the exponential decay of $P(t,k)$ at late times, such that its zero gives the expected value for the activity per unit time in the NESS. The probability of observing trajectories with activity smaller than average decays at a rate independent of $N$, as the inactive regime is dominated by the boundary driving. In appendix \ref{SCGFabsorbing} we display similar plots for a value of $\lambda = 0.7 < \lambda_c$ in the absorbing phase.   

The susceptibility $\chi(s) = \partial_s^2 \Theta(s)$ is shown in the left panel of Fig.~\ref{fig:criticalsN}. It shows a peak at the critical value $s_c(N)$ where the SCGF kinks and the activity jumps. Similar to \cite{banuls2019using}, we find that the behaviour of the susceptibility for $s< s_c(N)$ is almost universal: leading into the peak, the susceptibility shows a scaling behavior $\chi \sim (-s)^{-\gamma}$ with exponent $\gamma = 1.4498(6)$ regardless of system size. This exponent does depend on the value of $\lambda$. 

The location of $s_c(N)$ can be estimated by the location of the peak in $\chi(s)$. This reveals that $s_c(N)$ in turn satisfies a scaling law $s_c(N) \sim - N^{-\alpha}$ whenever $ \lambda > \lambda_c $. We show the critical values $s_c(N)$, as obtained from the peak in the susceptibility as a function of $1/N$ in the middle panel of Fig.~\ref{fig:criticalsN}. The dashed lines are linear fits used to obtain $\alpha$, which is shown against the infection parameter $\lambda $ in the right panel of Fig.~\ref{fig:criticalsN}. 

\subsection{Density of infected sites at finite $s$}\label{sec:density}

From the optimized tilted probability vector $\ket{\Psi_0(s)}$ we can obtain information on the density of infected sites at finite values of the dual parameter $s$. This gives the expectation value for the density on trajectories weighted by their activity; for $s<0$, inactive trajectories are more likely, while $s>0$ corresponds to more active trajectories. 

\begin{figure*}[t]
	\includegraphics[height=0.23\textwidth]{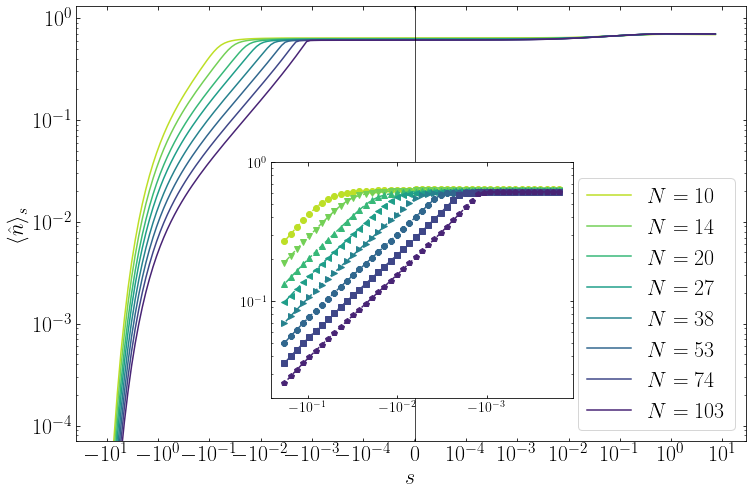}
	\includegraphics[height=0.23\textwidth]{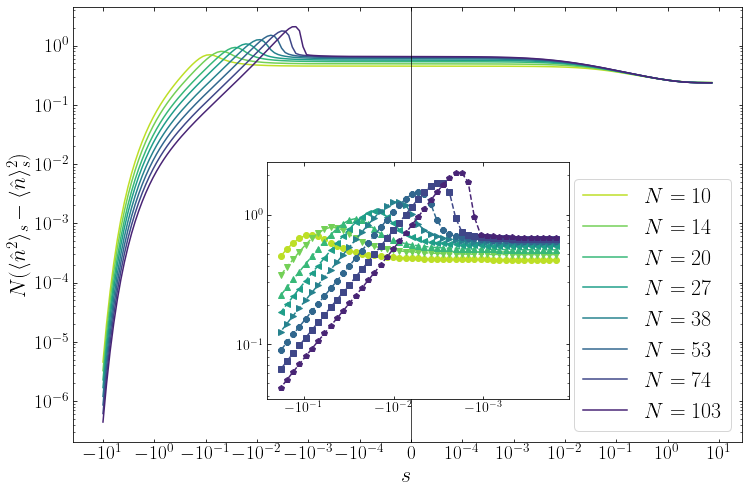}
	\includegraphics[height=0.23\textwidth]{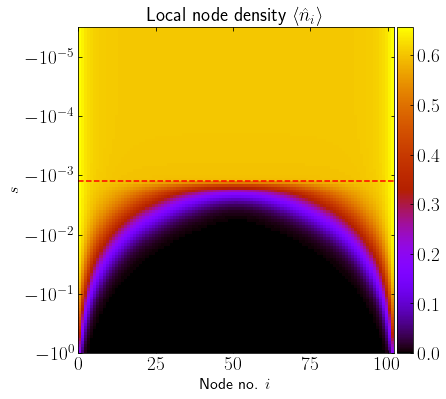}
	\caption{\label{fig:density} The left panel shows the density of infected sites in the chain of various lengths at $\lambda= 2$, as a function of $s$. Trajectories with lower activity clearly correspond to those with less dense configurations. The middle panel displays the variance in global density, which peaks when the expected density shows a sudden drop. The right panel show the local node density as a function of $s$ and node position in the chain for $N=103$ respectively. The dashed red line indicates the critical value $s_c(N=103)$ as determined from the peak in susceptibility.}
\end{figure*}

The density in the chain is computed efficiently by contracting an MPO as \eqref{densexp}, now at finite dual parameter $s$. Likewise, the variance in density is computed by contracting the square of the density MPO with the obtained tilted probability vector. 
In Fig.~\ref{fig:density}, we show the density of infected sites and the variance in density $N \left( \vev{\hat{n}^2 } - \vev{\hat{n}}^2\right)$ in the chain at various lengths as a function of the tilting parameter $s$ at $\lambda =2 > \lambda_c$. It is clear from the plot that less active trajectories (lower $s<0$) correspond to less dense configurations, while more active trajectories are hardly any denser than those of the Markovian processes at $s=0$. At the critical points $s_c(N)$, the expected density drops abruptly and the variance in density shows a peak. The heat plot in the right panel of Fig.~\ref{fig:density} shows the density per site in the chain for $N=103$. For $s > s_c$, the density is uniform over the whole chain. The active-inactive transition at $s_c$ (indicated by the red dashed line) clearly corresponds to a sharp drop in density in the middle of the chain. Below $s_c$, one can observe the diffusion from the boundaries due to the driving terms \eqref{Wdriv}: only the region close to the boundary stays infected.


\section{Entropic measures}\label{sec:entropic}

The formulation of the tilted many-body probability distribution as an MPS makes it possible to efficiently compute a variety of entropic measures for the stochastic process. As we are optimizing for a probability distribution, the Shannon entropy $H(\mathbf{x})$ would be the first quantity to look at, followed closely by the mutual information between two halves of the chain. However, computing $H(\mathbf{x}) = - \sum_\mathbf{x} P(\mathbf{x}) \log P(\mathbf{x})$ over all configurations $\mathbf{x}$ would require taking the logarithm of an MPS, for which there are no efficient methods known to us at the moment. Fortunately, entropic measures based on (second) R\'enyi entropy can be computed efficiently \cite{wolf2008area,scalet2021computable}. 

The \emph{second R\'enyi entropy}, or the \emph{Collision entropy} $H_2(\mathbf{x})$ is defined as:
\begin{equation}
	H_2(\mathbf{x}) = - \log_2 \sum_\mathbf{x} P(\mathbf{x})^2 \,.
\end{equation}
This is obtained efficiently as the $L^2$-norm of the distribution $\ket{\Psi_0(s)}$, as we can express the collision entropy as:
\begin{equation}\label{secrenen}
	H_2(\mathbf{x}) = - \log_2 \left( \frac{\bracket{\Psi_0}{\Psi_0}}{\bracket{\textbf{1}}{\Psi_0}^2 } \right)\,.
\end{equation}
Here $\bracket{\Psi_0}{\Psi_0}$ is computed efficiently by contracting the following tensor network horizontally
\begin{equation}
	\bracket{\Psi_0}{\Psi_0} ={\scriptsize
		\begin{array}{c}
		\begin{tikzpicture}
		\node[shape=circle,draw=black,fill=orange!20] (A1) at (0,0) {$A_1$};
		\node[shape=circle,draw=black,fill=orange!20] (A2) at (1,0) {$A_2$};
		\node[shape=circle] (dots) at (2,0) {$\ldots$};
		\node[shape=circle] (gdots) at (2,-1) {$\ldots$};
		\node[shape=circle,draw=black,fill=orange!20] (AN) at (3,0) {$A_N$};
		\node[shape=circle,draw=black,fill=orange!20] (g1) at (0,-1) {$A_1$};
		\node[shape=circle,draw=black,fill=orange!20] (g2) at (1,-1) {$A_2$};
		\node[shape=circle,draw=black,fill=orange!20] (gN) at (3,-1) {$A_N$};
		\path [-,line width=2pt] (A1) edge node[above] {} (A2);
		\path [-,line width=2pt] (A2) edge node[above] {} (dots);
		\path [-,line width=2pt] (AN) edge node[above] {} (dots);
		\path [-,line width=2pt] (g1) edge node[above] {} (g2);
		\path [-,line width=2pt] (g2) edge node[above] {} (gdots);
		\path [-,line width=2pt] (gN) edge node[above] {} (gdots);
		\path [-] (A1) edge node[left] {} (g1);
		\path [-] (A2) edge node[left] {} (g2);
		\path [-] (AN) edge node[left] {} (gN);
		\end{tikzpicture}
		\end{array} }\,.
\end{equation}

Considering a partition of the chain in two 
regions, $L$~and~$R$, the \emph{second R\'enyi entanglement entropy} $H_2(L)$ of the region $L$ is defined as the second R\'enyi entropy of the states distribution marginalised over the complementary region $R$. 
In terms of tensor networks, 
this can be computed by first contracting all sites in region $R$ with the local flat state
$\bra{1} = \begin{pmatrix} 1& 1 \end{pmatrix}$, 
followed by contracting the marginal distribution over $L$ with a copy of itself. Hence, we have:
\begin{equation}
    H_2(L) = - \log_2 \left( \frac{ \bracket{\Psi_0(L)}{\Psi_0(L)}}{\bracket{\textbf{1}}{\Psi_0}^2 } \right)
\end{equation}
with
\begin{equation}
    \bracket{\Psi_0(L)}{\Psi_0(L)} = {\scriptsize
        \begin{array}{c}
            \begin{tikzpicture}
                \node[shape=circle,draw=black,fill=orange!20] (A1) at (0,0) {$A_1$};
                \node[shape=circle] (dots) at (1,0) {$\ldots$};
                \node[shape=circle] (gdots) at (1,-1) {$\ldots$};
                \node[shape=circle] (dots2) at (4,0) {$\ldots$};
                \node[shape=circle] (gdots2) at (4,-1) {$\ldots$};
                \node[shape=circle,draw=black,fill=green!20] (normb) at (3,1) {};
                \node[shape=circle,draw=black,fill=green!20] (normb2) at (3,-2) {};
                \node[shape=circle,draw=black,fill=green!20] (normN) at (5,1) {};
                \node[shape=circle,draw=black,fill=green!20] (normN2) at (5,-2) {};
                \node[shape=circle,draw=black,fill=orange!20] (Aa) at (2,0) {$A_L$};
                \node[shape=circle,draw=black,fill=orange!20] (ga) at (2,-1) {$A_L$};
                \node[shape=circle,draw=black,fill=orange!20] (Aa1) at (3,0) {$A_R$}; 
                \node[shape=circle,draw=black,fill=orange!20] (AN) at (5,0) {$A_N$};
                \node[shape=circle,draw=black,fill=orange!20] (g1) at (0,-1) {$A_1$};
                \node[shape=circle,draw=black,fill=orange!20] (ga1) at (3,-1) {$A_R$}; 
                \node[shape=circle,draw=black,fill=orange!20] (gN) at (5,-1) {$A_N$};
                \path [-,line width=2pt] (A1) edge node[above] {} (dots);
                \path [-,line width=2pt] (dots) edge node[above] {} (Aa);
                \path [-,line width=2pt] (Aa1) edge node[above] {} (Aa);
                \path [-,line width=2pt] (g1) edge node[above] {} (gdots);
                \path [-,line width=2pt] (gdots) edge node[above] {} (ga);
                \path [-,line width=2pt] (ga) edge node[above] {} (ga1);
                \path [-,line width=2pt] (ga1) edge node[above] {} (gdots2);
                \path [-,line width=2pt] (Aa1) edge node[above] {} (dots2);
                \path [-,line width=2pt] (dots2) edge node[above] {} (AN);
                \path [-,line width=2pt] (gdots2) edge node[above] {} (gN);
                \path [-] (A1) edge node[left] {} (g1);
                \path [-] (Aa) edge node[left] {} (ga);
                \path [-] (Aa1) edge node[left] {} (normb);
                \path [-] (ga1) edge node[left] {} (normb2);
                \path [-] (AN) edge node[left] {} (normN);
                \path [-] (gN) edge node[left] {} (normN2);
\draw [
    thick,
    decoration={
        brace,
        raise=0.5cm
    },
    decorate
] (A1) -- (Aa) 
node [pos=0.5,anchor=north,yshift=1cm] {$L$};
\draw [
    thick,
    decoration={
        brace,
        raise=0.5cm
    },
    decorate
] (Aa1) -- (AN) 
node [pos=0.5,anchor=north,yshift=1cm] {$R$}; 
            \end{tikzpicture}
    \end{array} }\,.
\end{equation}
Here the green nodes denote local flat states $\bra{\mathbf{1}} $. The second R\'enyi entanglement entropy $H_2(R)$ for region $R$, defined as the complement of $L$, is computed similarly, by marginalizing over the sites in region $L$ followed by computing the second R\'enyi entropy of region $R$.

\begin{figure*}[!ht]
	\includegraphics[width=0.9\textwidth]{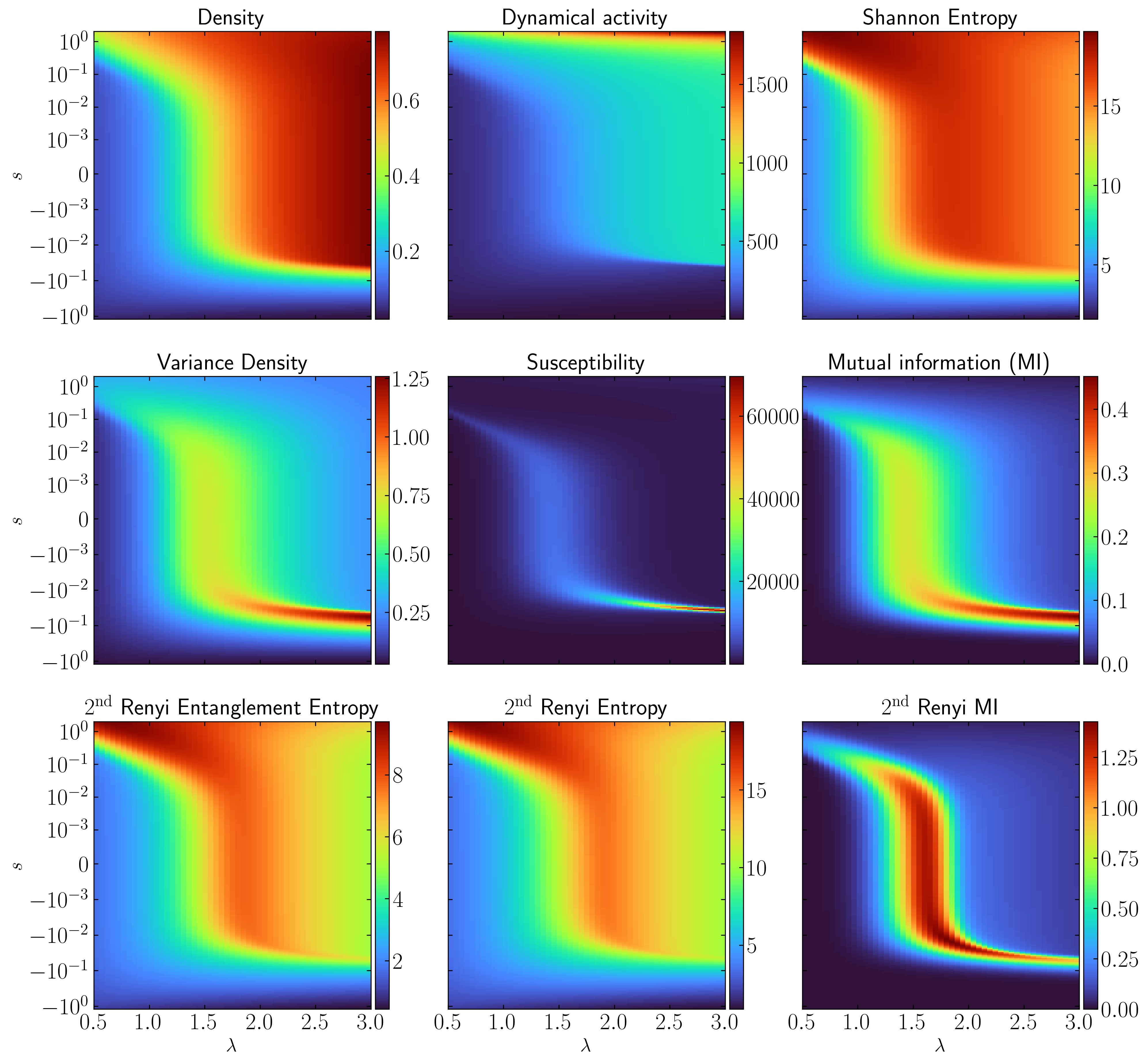}
	\caption{\label{fig:heatplots} Various observables for the contact process in the chain of length $N=20$ in the $(\lambda,s)$ plane.}
\end{figure*}

Using the above quantities, one can compute the \emph{second R\'enyi mutual information} (RMI) $H_2(L;R)$ between the two subregions $L$ and $R$:
\begin{equation}\label{secRenMI}
    H_2(L;R) = H_2(L) +  H_2(R) - H_2(L,R) \,.
\end{equation}
Here $H_2(L,R) = H_2(\mathbf{x})$ is defined above in \eqref{secrenen} and $H_2(L)$ and $H_2(R)$ are the second R\'enyi entanglement entropies of regions $L$ and $R$, respectively.

The RMI is particularly interesting since it provides a measure of the amount of information contained in a sub-region of the system about the rest of the system. Conventional measures based on the correlation length are susceptible to overlooking hidden correlations, whereas the RMI is based on the total amount of information of a sub-system about the complementing region, without overlooking hidden correlations \cite{wolf2008area}. Furthermore, it has been shown in \cite{iaconis2013detecting} that the second RMI provides a way to detect phase transitions in classical equilibrium systems, without any prior knowledge of an order parameter or other thermodynamic quantities. Below, we demonstrate that the second RMI can be computed efficiently from the MPS representation of the NESS for the contact process, and use it to identify the absorbing state transition by a finite-size scaling argument.  

In Fig.~\ref{fig:heatplots}, we show and compare various quantities in the $(\lambda, s)$-plane for chain of length $N=20$. For this chain length, it is still computationally feasible to compute the complete Shannon entropy and mutual information between two halves of the chain. These plots have been added for comparison. The division into subregions in this plot is always taken to be the half chain and driving at both boundaries was used here.

We find that the second R\'enyi mutual information shows a clear peak close to the region where the susceptibility and mutual information is maximal. However, a closer inspection shows that the peak in $H_2(L;R)$ is shifted slightly towards higher values of $\lambda$. This shows that the second R\'enyi mutual information is maximal just above the critical transmission rate $\lambda_c$. 

\begin{figure*}[!ht]
	\includegraphics[height=0.3\textwidth]{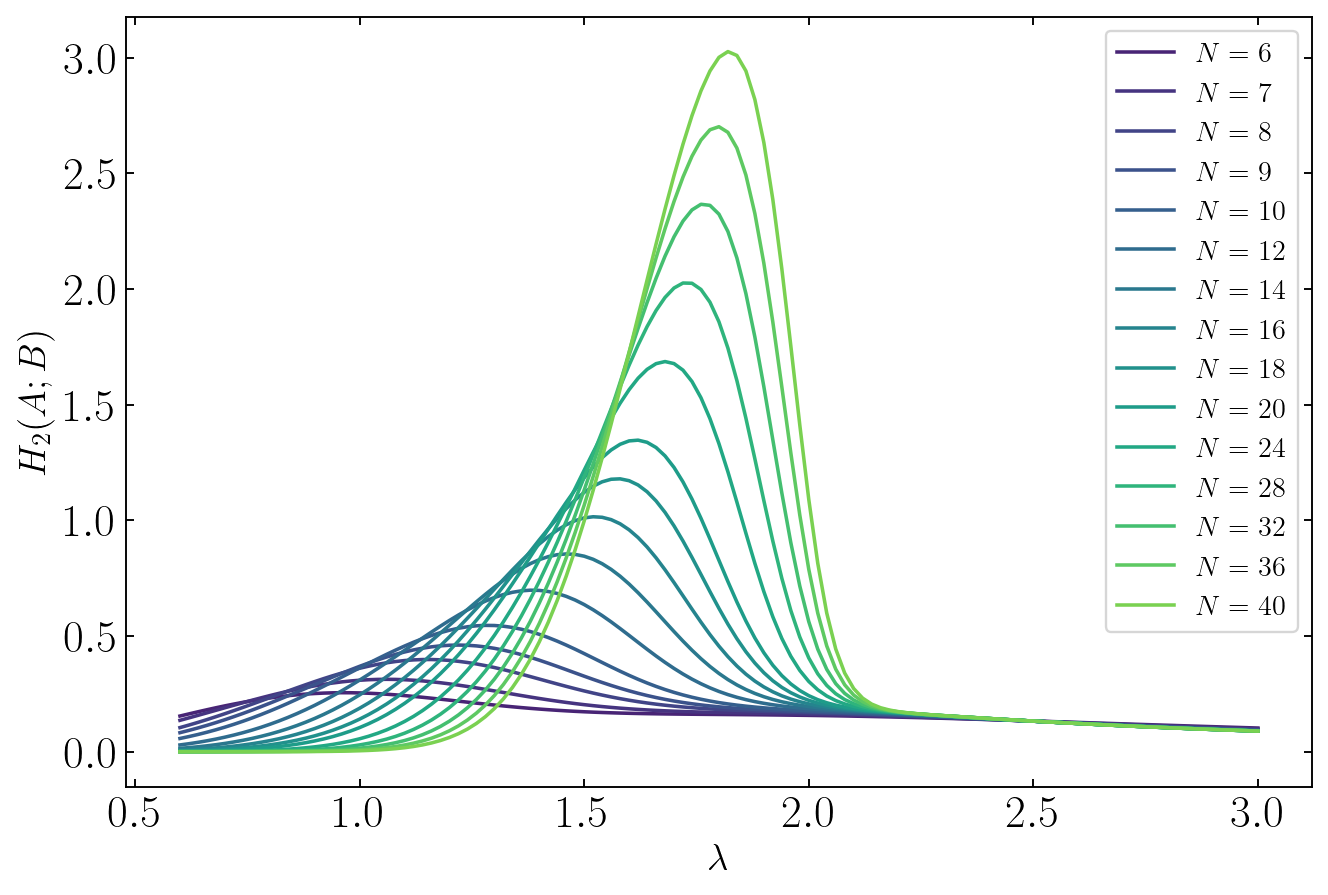}
	\includegraphics[height=0.3\textwidth]{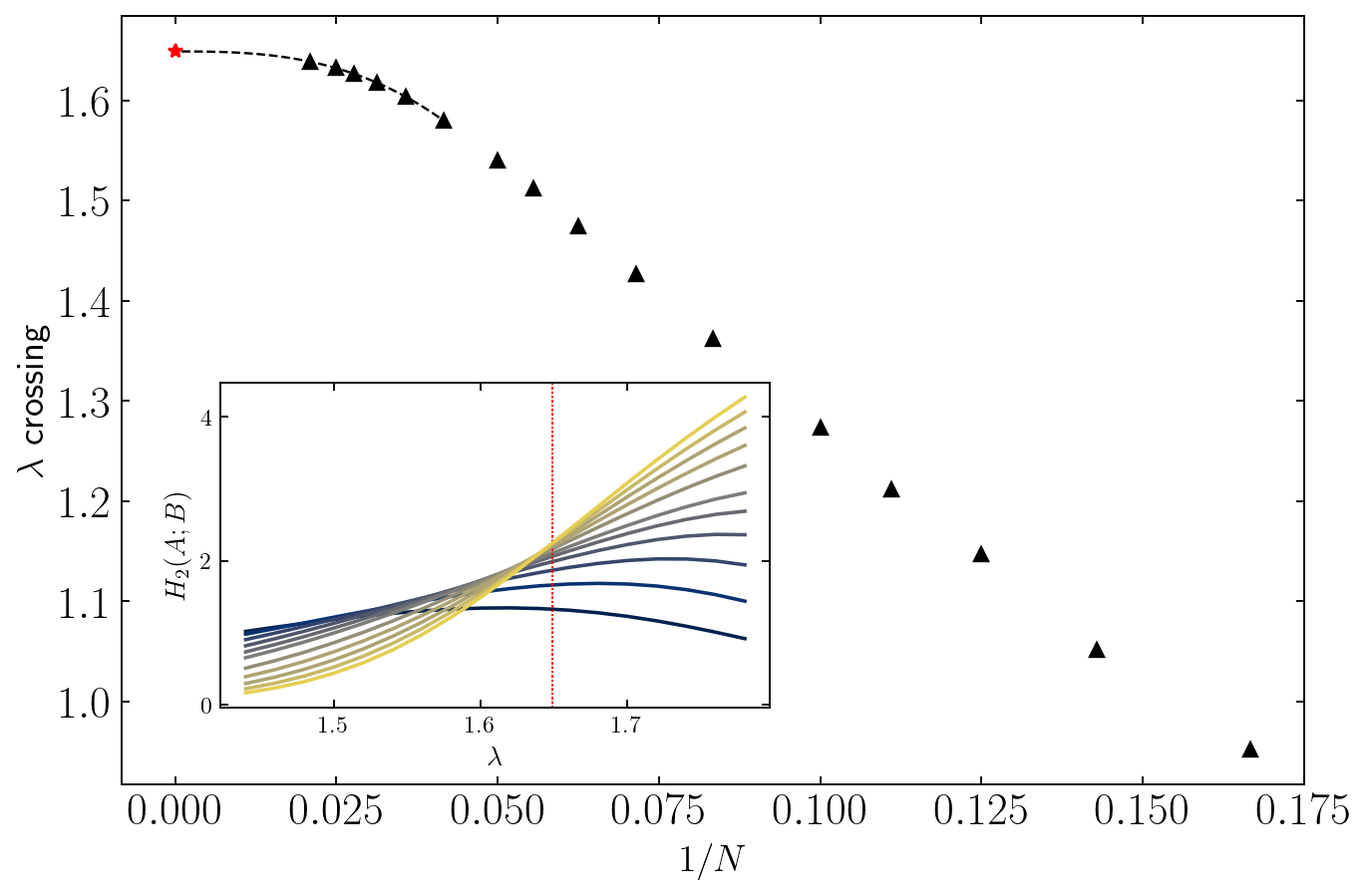}
	\caption{\label{fig:RMI} The second R\'enyi mutual information (RMI) (left) in the NESS as a function of $\lambda$ (with boundary driving). The right panel shows the locations $\lambda(N)$ where the RMI curves for $N$ and $2N$ cross, as a function of $1/N$. The red star marks the critical value of the 1D contact process $\lambda_c = 1.6489$. The inset shows a detail of the crossing points in the RMI with $N = 20, 24, 28, 32 ,36, 40, 48, 56, 64, 72, 80$. Dotted line is an extrapolation to $N \to \infty$ based on Eq.~\eqref{scalingfunction}. }
\end{figure*}

In \cite{iaconis2013detecting}, the second RMI was used to detect phase transitions in classical equilibrium models. The analysis relied on a scaling argument, which shows the RMI contains a constant contribution (i.e. independent of sub-system area) which changes sign across the critical temperature $T_c$. This causes the finite-size RMI curves to ``fan out'' at $T_c$ for different system sizes, producing a crossing of the curves. In our case, our system is driven out-of-equilibrium, however, the RMI curves still show a crossing close to the critical infection rate $\lambda_c$. This is demonstrated in Fig.~\ref{fig:RMI}, which displays the RMI in the NESS for various system sizes as a function of $\lambda$. The right panel of Fig. \ref{fig:RMI} displays the crossing points of the RMI at system sizes $N$ and $2N$ for increasing $N$. For small system sizes, there is a considerable finite-size effect, as the intersections are not occuring at the same locations. As $N$ increases, the location of the intersection creeps closer to $\lambda_c$. For chains of length $\sim 80$, the crossings already occur very close to the critical infection rate. 

We extrapolate the data for the largest available values of $N$ using the scaling Ansatz
\begin{equation}\label{scalingfunction}
    \lambda_c(N) = \lambda_c \left( 1 + \frac{b}{N^\chi} + \ldots \right) \,.
\end{equation}
This allows us to derive a critical transmission rate $\lambda_c = 1.6487(6)$, independently from using any order parameter. The obtained value is consistent with the known critical value for $\lambda$. Additionally, we obtain a value for the sub-leading scaling with $N$: $\chi = 2.77(5)$. This analysis demonstrates the ability to detect dynamical phase transitions on the basis of the RMI alone, without the need for an order parameter.




\section{Conclusions and discussion}\label{sec:conclusions}

In this paper, we have used the Matrix Product State to study the one-dimensional contact process. 
We have found that efficient representations for the many-body probability vector in the thermodynamic limit $t \to \infty$ are obtained in the active and inactive regimes, while at the critical threshold for the active/inactive phase transition, the bond dimension of the MPS grows at most linear with system size. 
We have illustrated how the MPS can be used for efficient computation of the distribution of gaps in chain, even for large gap sizes corresponding to extremely rare events.
We used the MPS to obtain accurate approximations of the scaled cumulant generating function for the dynamical activity, which is dual to the large deviation rate function by Legendre transform. This analysis sheds light on how trajectories with different dynamical activity contribute to the properties of the system in the late time regime.

Our approach enables one to study entropic measures in a straightforward way, which is challenging to do with other approaches. In particular, we compute the second Renyi mutual information for the contact process. Previous methods obtained the second RMI for equilibrium models relying on Monte Carlo simulations of the replicated partition function (see for instance \cite{iaconis2013detecting}). In our case, we are lacking a Boltzmann distribution as we are looking at an out-of-equilibrium problem, so it is not clear how to obtain the RMI from applying the replica trick on a partition function. Instead, we obtained the second RMI directly from the MPS representation of the NESS and demonstrated its use to detect dynamical phase transitions.

This work opens the doors to tensor network applications in mathematical epidemiology. One could imagine formulating a tensor network state for more involved compartmental models of epidemiology along the lines illustrated in this paper, but with a closer connection to real-world infectious disease outbreaks. The tensor network could then be used to obtain probability distributions over observables of relevance to the tracking and forecasting of the epidemic state through a population of interest. Measurements on the populace, such as test and tracing policies could then be implemented into the tensor network by projecting certain sites on the measurement outcome. In this way, the state will become closer to the actual state of the population, while retaining the information of past correlations between sites not projected on. 

While the one-dimensional contact process remains an interesting system for statistical physicists, in mathematical epidemiology one is more often interested in time-dependent spreading processes on complex networks \cite{PastorSatorras2015}. It is therefore of great interest to explore how the methods described in this paper can be used to predict and forecast epidemic spreading in realistic scenarios. Major steps in this direction would be an extension to study the dynamical evolution by time-evolving block decimation algorithms (TEBD) \cite{vidal2004efficient} and to extend the methods to spreading processes on real-world, complex contact networks. We hope to address such questions in future work.


\begin{figure}[!th]
	\includegraphics[width=\columnwidth]{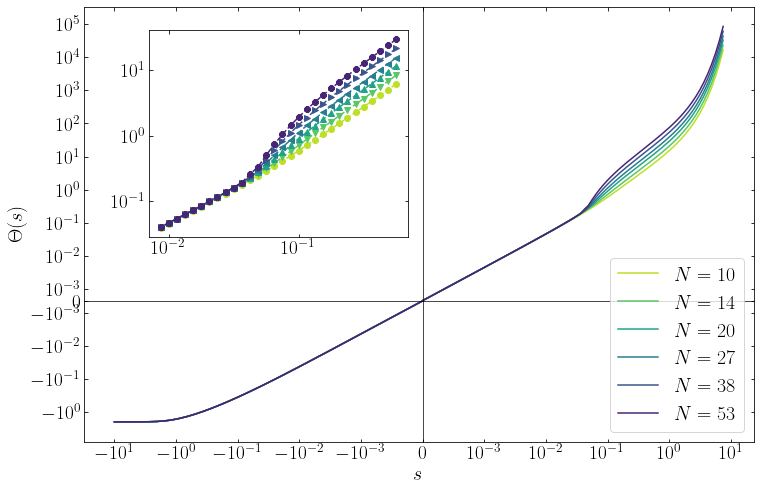}
	\includegraphics[width=\columnwidth]{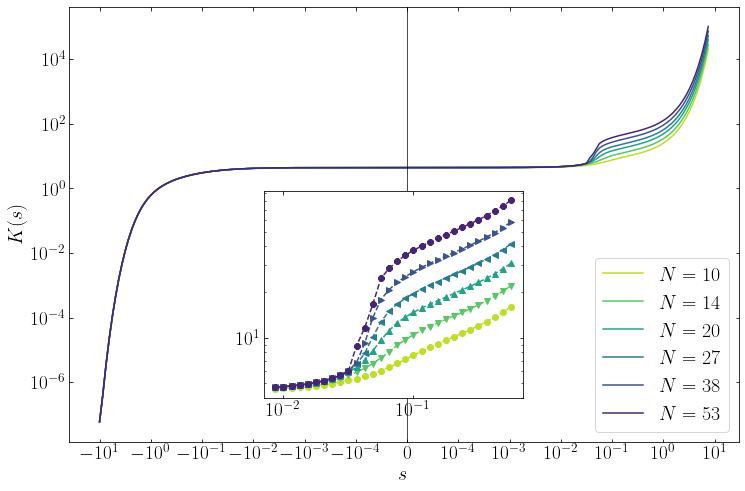}
	\includegraphics[width=\columnwidth]{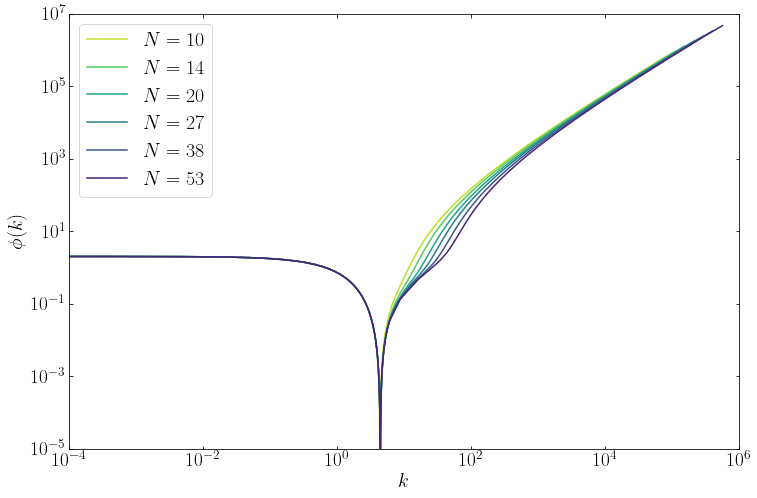}
	\caption{\label{fig:FEabsorbing} The scaled cumulant generating function (top) for various $N$ at $\lambda = 0.7 < \lambda_c$ using the boundary driving protocol. The middle panel shows the activity per unit time as a function of $s$, obtained as $K(s) = \partial_s \Theta(s)$. The rate function (bottom) shows zeros at the expected value for the activity. Due to the boundary driving, inactive trajectories (small $k$) all have the same constant decay rate $\phi(k)$.}
\end{figure}

\begin{acknowledgments}
	This project has received funding from the European Research Council (ERC) under the European Union’s Horizon 2020 research and innovation programme (grant agreement No 101001604). WM acknowledges the support from the NWO Klein grant awarded to NWA route 2 in 2020.
\end{acknowledgments}

\appendix
\section{Scaled cumulant generating function for $\lambda < \lambda_c$}
\label{SCGFabsorbing}

Here we display our results for the SCGF $\Theta(s)$ for values of $\lambda$ in the absorbing phase. Figure~\ref{fig:FEabsorbing} displays the SCGF $\Theta(s)$ for $\lambda = 0.7 < \lambda_c$. Note that we plot here the SCGF and not the SCGF per site, as was plotted in Fig.~\ref{fig:FEresults} for $\lambda > \lambda_c$. We see that in this case for any $N$, the SCFG is approximately linear around $s=0$ with the same $N$ independent slope. This implies that the activity is independent of $N$, as illustrated as well by the middle panel of Fig.~\ref{fig:FEabsorbing}. Only for trajectories tilted towards the active region ($s>0$) a jump in the dynamical activity is visible. 

The lower panel of Fig.~\ref{fig:FEresults} displays the rate function for the dynamical activity when $\lambda = 0.7 < \lambda_c$. The plot shows that the likelihood of observing trajectories with a dynamical activity which is average or lower decay exponentially with a rate independent of $N$, which is an effect of the boundary driving term. For trajectories with activity larger than average, the decay rate does depend on $N$, such that larger $N$ corresponds to a lower rate function.



\bibliography{biblio}

\begin{thebibliography}{77}%
\makeatletter
\providecommand \@ifxundefined [1]{%
 \@ifx{#1\undefined}
}%
\providecommand \@ifnum [1]{%
 \ifnum #1\expandafter \@firstoftwo
 \else \expandafter \@secondoftwo
 \fi
}%
\providecommand \@ifx [1]{%
 \ifx #1\expandafter \@firstoftwo
 \else \expandafter \@secondoftwo
 \fi
}%
\providecommand \natexlab [1]{#1}%
\providecommand \enquote  [1]{``#1''}%
\providecommand \bibnamefont  [1]{#1}%
\providecommand \bibfnamefont [1]{#1}%
\providecommand \citenamefont [1]{#1}%
\providecommand \href@noop [0]{\@secondoftwo}%
\providecommand \href [0]{\begingroup \@sanitize@url \@href}%
\providecommand \@href[1]{\@@startlink{#1}\@@href}%
\providecommand \@@href[1]{\endgroup#1\@@endlink}%
\providecommand \@sanitize@url [0]{\catcode `\\12\catcode `\$12\catcode
  `\&12\catcode `\#12\catcode `\^12\catcode `\_12\catcode `\%12\relax}%
\providecommand \@@startlink[1]{}%
\providecommand \@@endlink[0]{}%
\providecommand \url  [0]{\begingroup\@sanitize@url \@url }%
\providecommand \@url [1]{\endgroup\@href {#1}{\urlprefix }}%
\providecommand \urlprefix  [0]{URL }%
\providecommand \Eprint [0]{\href }%
\providecommand \doibase [0]{https://doi.org/}%
\providecommand \selectlanguage [0]{\@gobble}%
\providecommand \bibinfo  [0]{\@secondoftwo}%
\providecommand \bibfield  [0]{\@secondoftwo}%
\providecommand \translation [1]{[#1]}%
\providecommand \BibitemOpen [0]{}%
\providecommand \bibitemStop [0]{}%
\providecommand \bibitemNoStop [0]{.\EOS\space}%
\providecommand \EOS [0]{\spacefactor3000\relax}%
\providecommand \BibitemShut  [1]{\csname bibitem#1\endcsname}%
\let\auto@bib@innerbib\@empty
\bibitem [{\citenamefont {Lloyd}\ and\ \citenamefont
  {May}(2001)}]{lloyd2001viruses}%
  \BibitemOpen
  \bibfield  {author} {\bibinfo {author} {\bibfnamefont {A.~L.}\ \bibnamefont
  {Lloyd}}\ and\ \bibinfo {author} {\bibfnamefont {R.~M.}\ \bibnamefont
  {May}},\ }\bibfield  {title} {\bibinfo {title} {How viruses spread among
  computers and people},\ }\href@noop {} {\bibfield  {journal} {\bibinfo
  {journal} {Science}\ }\textbf {\bibinfo {volume} {292}},\ \bibinfo {pages}
  {1316} (\bibinfo {year} {2001})}\BibitemShut {NoStop}%
\bibitem [{\citenamefont {Albert}\ and\ \citenamefont
  {Barabási}(2002)}]{albert2002statistical}%
  \BibitemOpen
  \bibfield  {author} {\bibinfo {author} {\bibfnamefont {R.}~\bibnamefont
  {Albert}}\ and\ \bibinfo {author} {\bibfnamefont {A.-L.}\ \bibnamefont
  {Barabási}},\ }\bibfield  {title} {\bibinfo {title} {Statistical mechanics
  of complex networks},\ }\href {https://doi.org/10.1103/RevModPhys.74.47}
  {\bibfield  {journal} {\bibinfo  {journal} {Reviews of modern physics}\
  }\textbf {\bibinfo {volume} {74}},\ \bibinfo {pages} {47} (\bibinfo {year}
  {2002})}\BibitemShut {NoStop}%
\bibitem [{\citenamefont {Boccaletti}\ \emph {et~al.}(2006)\citenamefont
  {Boccaletti}, \citenamefont {Latora}, \citenamefont {Moreno}, \citenamefont
  {Chavez},\ and\ \citenamefont {Hwang}}]{boccaletti2006complex}%
  \BibitemOpen
  \bibfield  {author} {\bibinfo {author} {\bibfnamefont {S.}~\bibnamefont
  {Boccaletti}}, \bibinfo {author} {\bibfnamefont {V.}~\bibnamefont {Latora}},
  \bibinfo {author} {\bibfnamefont {Y.}~\bibnamefont {Moreno}}, \bibinfo
  {author} {\bibfnamefont {M.}~\bibnamefont {Chavez}},\ and\ \bibinfo {author}
  {\bibfnamefont {D.-U.}\ \bibnamefont {Hwang}},\ }\bibfield  {title} {\bibinfo
  {title} {Complex networks: Structure and dynamics},\ }\href
  {https://doi.org/10.1016/j.physrep.2005.10.009} {\bibfield  {journal}
  {\bibinfo  {journal} {Physics reports}\ }\textbf {\bibinfo {volume} {424}},\
  \bibinfo {pages} {175–308} (\bibinfo {year} {2006})}\BibitemShut {NoStop}%
\bibitem [{\citenamefont {Castellano}\ \emph {et~al.}(2009)\citenamefont
  {Castellano}, \citenamefont {Fortunato},\ and\ \citenamefont
  {Loreto}}]{castellano2009statistical}%
  \BibitemOpen
  \bibfield  {author} {\bibinfo {author} {\bibfnamefont {C.}~\bibnamefont
  {Castellano}}, \bibinfo {author} {\bibfnamefont {S.}~\bibnamefont
  {Fortunato}},\ and\ \bibinfo {author} {\bibfnamefont {V.}~\bibnamefont
  {Loreto}},\ }\bibfield  {title} {\bibinfo {title} {Statistical physics of
  social dynamics},\ }\href {https://doi.org/10.1103/RevModPhys.81.591}
  {\bibfield  {journal} {\bibinfo  {journal} {Reviews of modern physics}\
  }\textbf {\bibinfo {volume} {81}},\ \bibinfo {pages} {591} (\bibinfo {year}
  {2009})}\BibitemShut {NoStop}%
\bibitem [{\citenamefont {Vespignani}(2012)}]{vespignani2012modelling}%
  \BibitemOpen
  \bibfield  {author} {\bibinfo {author} {\bibfnamefont {A.}~\bibnamefont
  {Vespignani}},\ }\bibfield  {title} {\bibinfo {title} {Modelling dynamical
  processes in complex socio-technical systems},\ }\href@noop {} {\bibfield
  {journal} {\bibinfo  {journal} {Nature physics}\ }\textbf {\bibinfo {volume}
  {8}},\ \bibinfo {pages} {32} (\bibinfo {year} {2012})}\BibitemShut {NoStop}%
\bibitem [{\citenamefont {Artime}\ and\ \citenamefont
  {De~Domenico}(2022)}]{artime2022origin}%
  \BibitemOpen
  \bibfield  {author} {\bibinfo {author} {\bibfnamefont {O.}~\bibnamefont
  {Artime}}\ and\ \bibinfo {author} {\bibfnamefont {M.}~\bibnamefont
  {De~Domenico}},\ }\href@noop {} {\bibinfo {title} {From the origin of life to
  pandemics: emergent phenomena in complex systems}} (\bibinfo {year}
  {2022})\BibitemShut {NoStop}%
\bibitem [{\citenamefont {Gleeson}(2013)}]{gleeson2013binary}%
  \BibitemOpen
  \bibfield  {author} {\bibinfo {author} {\bibfnamefont {J.~P.}\ \bibnamefont
  {Gleeson}},\ }\bibfield  {title} {\bibinfo {title} {Binary-state dynamics on
  complex networks: Pair approximation and beyond},\ }\href
  {https://doi.org/10.1103/PhysRevX.3.021004} {\bibfield  {journal} {\bibinfo
  {journal} {Physical Review X}\ }\textbf {\bibinfo {volume} {3}},\ \bibinfo
  {pages} {021004} (\bibinfo {year} {2013})}\BibitemShut {NoStop}%
\bibitem [{\citenamefont {Masuda}\ \emph {et~al.}(2017)\citenamefont {Masuda},
  \citenamefont {Porter},\ and\ \citenamefont {Lambiotte}}]{masuda2017random}%
  \BibitemOpen
  \bibfield  {author} {\bibinfo {author} {\bibfnamefont {N.}~\bibnamefont
  {Masuda}}, \bibinfo {author} {\bibfnamefont {M.~A.}\ \bibnamefont {Porter}},\
  and\ \bibinfo {author} {\bibfnamefont {R.}~\bibnamefont {Lambiotte}},\
  }\bibfield  {title} {\bibinfo {title} {Random walks and diffusion on
  networks},\ }\href {https://doi.org/10.1016/j.physrep.2017.07.007} {\bibfield
   {journal} {\bibinfo  {journal} {Physics reports}\ }\textbf {\bibinfo
  {volume} {716}},\ \bibinfo {pages} {1} (\bibinfo {year} {2017})}\BibitemShut
  {NoStop}%
\bibitem [{\citenamefont {Ghavasieh}\ \emph {et~al.}(2020)\citenamefont
  {Ghavasieh}, \citenamefont {Nicolini},\ and\ \citenamefont
  {De~Domenico}}]{ghavasieh2020statistical}%
  \BibitemOpen
  \bibfield  {author} {\bibinfo {author} {\bibfnamefont {A.}~\bibnamefont
  {Ghavasieh}}, \bibinfo {author} {\bibfnamefont {C.}~\bibnamefont
  {Nicolini}},\ and\ \bibinfo {author} {\bibfnamefont {M.}~\bibnamefont
  {De~Domenico}},\ }\bibfield  {title} {\bibinfo {title} {Statistical physics
  of complex information dynamics},\ }\href
  {https://doi.org/10.1103/PhysRevE.102.052304} {\bibfield  {journal} {\bibinfo
   {journal} {Physical Review E}\ }\textbf {\bibinfo {volume} {102}},\ \bibinfo
  {pages} {052304} (\bibinfo {year} {2020})}\BibitemShut {NoStop}%
\bibitem [{\citenamefont {Merbis}\ and\ \citenamefont
  {de~Domenico}(2023{\natexlab{a}})}]{merbis2023emergent}%
  \BibitemOpen
  \bibfield  {author} {\bibinfo {author} {\bibfnamefont {W.}~\bibnamefont
  {Merbis}}\ and\ \bibinfo {author} {\bibfnamefont {M.}~\bibnamefont
  {de~Domenico}},\ }\href@noop {} {\bibinfo {title} {Emergent information
  dynamics in many-body interconnected systems}} (\bibinfo {year}
  {2023}{\natexlab{a}}),\ \Eprint {https://arxiv.org/abs/2305.05428}
  {arXiv:2305.05428 [physics.soc-ph]} \BibitemShut {NoStop}%
\bibitem [{\citenamefont {Privman}(1990)}]{privman1990finite}%
  \BibitemOpen
  \bibfield  {author} {\bibinfo {author} {\bibfnamefont {V.}~\bibnamefont
  {Privman}},\ }\href@noop {} {\emph {\bibinfo {title} {Finite size scaling and
  numerical simulation of statistical systems}}}\ (\bibinfo  {publisher} {World
  Scientific},\ \bibinfo {year} {1990})\BibitemShut {NoStop}%
\bibitem [{\citenamefont {Gleeson}\ \emph {et~al.}(2012)\citenamefont
  {Gleeson}, \citenamefont {Melnik}, \citenamefont {Ward}, \citenamefont
  {Porter},\ and\ \citenamefont {Mucha}}]{gleeson2012accuracy}%
  \BibitemOpen
  \bibfield  {author} {\bibinfo {author} {\bibfnamefont {J.~P.}\ \bibnamefont
  {Gleeson}}, \bibinfo {author} {\bibfnamefont {S.}~\bibnamefont {Melnik}},
  \bibinfo {author} {\bibfnamefont {J.~A.}\ \bibnamefont {Ward}}, \bibinfo
  {author} {\bibfnamefont {M.~A.}\ \bibnamefont {Porter}},\ and\ \bibinfo
  {author} {\bibfnamefont {P.~J.}\ \bibnamefont {Mucha}},\ }\bibfield  {title}
  {\bibinfo {title} {Accuracy of mean-field theory for dynamics on real-world
  networks},\ }\href {https://doi.org/10.1103/PhysRevE.85.026106} {\bibfield
  {journal} {\bibinfo  {journal} {Physical Review E}\ }\textbf {\bibinfo
  {volume} {85}},\ \bibinfo {pages} {026106} (\bibinfo {year}
  {2012})}\BibitemShut {NoStop}%
\bibitem [{\citenamefont {Gao}\ \emph {et~al.}(2016)\citenamefont {Gao},
  \citenamefont {Barzel},\ and\ \citenamefont
  {Barab{\'a}si}}]{gao2016universal}%
  \BibitemOpen
  \bibfield  {author} {\bibinfo {author} {\bibfnamefont {J.}~\bibnamefont
  {Gao}}, \bibinfo {author} {\bibfnamefont {B.}~\bibnamefont {Barzel}},\ and\
  \bibinfo {author} {\bibfnamefont {A.-L.}\ \bibnamefont {Barab{\'a}si}},\
  }\bibfield  {title} {\bibinfo {title} {Universal resilience patterns in
  complex networks},\ }\href {https://doi.org/10.1038/nature16948} {\bibfield
  {journal} {\bibinfo  {journal} {Nature}\ }\textbf {\bibinfo {volume} {530}},\
  \bibinfo {pages} {307} (\bibinfo {year} {2016})}\BibitemShut {NoStop}%
\bibitem [{\citenamefont {Kiss}\ \emph {et~al.}(2017)\citenamefont {Kiss},
  \citenamefont {Miller}, \citenamefont {Simon} \emph
  {et~al.}}]{kiss2017mathematics}%
  \BibitemOpen
  \bibfield  {author} {\bibinfo {author} {\bibfnamefont {I.~Z.}\ \bibnamefont
  {Kiss}}, \bibinfo {author} {\bibfnamefont {J.~C.}\ \bibnamefont {Miller}},
  \bibinfo {author} {\bibfnamefont {P.~L.}\ \bibnamefont {Simon}}, \emph
  {et~al.},\ }\href@noop {} {\emph {\bibinfo {title} {Mathematics of epidemics
  on networks}}},\ Vol.\ \bibinfo {volume} {598}\ (\bibinfo  {publisher}
  {Springer},\ \bibinfo {year} {2017})\BibitemShut {NoStop}%
\bibitem [{\citenamefont {Young}\ \emph {et~al.}(2001)\citenamefont {Young},
  \citenamefont {Roberts},\ and\ \citenamefont
  {Stuhne}}]{young2001reproductive}%
  \BibitemOpen
  \bibfield  {author} {\bibinfo {author} {\bibfnamefont {W.~R.}\ \bibnamefont
  {Young}}, \bibinfo {author} {\bibfnamefont {A.~J.}\ \bibnamefont {Roberts}},\
  and\ \bibinfo {author} {\bibfnamefont {G.}~\bibnamefont {Stuhne}},\
  }\bibfield  {title} {\bibinfo {title} {Reproductive pair correlations and the
  clustering of organisms},\ }\href@noop {} {\bibfield  {journal} {\bibinfo
  {journal} {Nature}\ }\textbf {\bibinfo {volume} {412}},\ \bibinfo {pages}
  {328} (\bibinfo {year} {2001})}\BibitemShut {NoStop}%
\bibitem [{\citenamefont {Houchmandzadeh}(2009)}]{houchmandzadeh2009theory}%
  \BibitemOpen
  \bibfield  {author} {\bibinfo {author} {\bibfnamefont {B.}~\bibnamefont
  {Houchmandzadeh}},\ }\bibfield  {title} {\bibinfo {title} {Theory of neutral
  clustering for growing populations},\ }\href@noop {} {\bibfield  {journal}
  {\bibinfo  {journal} {Physical Review E}\ }\textbf {\bibinfo {volume} {80}},\
  \bibinfo {pages} {051920} (\bibinfo {year} {2009})}\BibitemShut {NoStop}%
\bibitem [{\citenamefont {Dumonteil}\ \emph {et~al.}(2014)\citenamefont
  {Dumonteil}, \citenamefont {Malvagi}, \citenamefont {Zoia}, \citenamefont
  {Mazzolo}, \citenamefont {Artusio}, \citenamefont {Dieudonn{\'e}},\ and\
  \citenamefont {De~Mulatier}}]{dumonteil2014particle}%
  \BibitemOpen
  \bibfield  {author} {\bibinfo {author} {\bibfnamefont {E.}~\bibnamefont
  {Dumonteil}}, \bibinfo {author} {\bibfnamefont {F.}~\bibnamefont {Malvagi}},
  \bibinfo {author} {\bibfnamefont {A.}~\bibnamefont {Zoia}}, \bibinfo {author}
  {\bibfnamefont {A.}~\bibnamefont {Mazzolo}}, \bibinfo {author} {\bibfnamefont
  {D.}~\bibnamefont {Artusio}}, \bibinfo {author} {\bibfnamefont
  {C.}~\bibnamefont {Dieudonn{\'e}}},\ and\ \bibinfo {author} {\bibfnamefont
  {C.}~\bibnamefont {De~Mulatier}},\ }\bibfield  {title} {\bibinfo {title}
  {Particle clustering in monte carlo criticality simulations},\ }\href@noop {}
  {\bibfield  {journal} {\bibinfo  {journal} {Annals of Nuclear Energy}\
  }\textbf {\bibinfo {volume} {63}},\ \bibinfo {pages} {612} (\bibinfo {year}
  {2014})}\BibitemShut {NoStop}%
\bibitem [{\citenamefont {Hethcote}(2000)}]{hethcote2000mathematics}%
  \BibitemOpen
  \bibfield  {author} {\bibinfo {author} {\bibfnamefont {H.~W.}\ \bibnamefont
  {Hethcote}},\ }\bibfield  {title} {\bibinfo {title} {The mathematics of
  infectious diseases},\ }\href {https://doi.org/10.1137/S0036144500371907}
  {\bibfield  {journal} {\bibinfo  {journal} {SIAM review}\ }\textbf {\bibinfo
  {volume} {42}},\ \bibinfo {pages} {599–653} (\bibinfo {year}
  {2000})}\BibitemShut {NoStop}%
\bibitem [{\citenamefont {Pastor-Satorras}\ and\ \citenamefont
  {Vespignani}(2001)}]{pastor2001epidemic}%
  \BibitemOpen
  \bibfield  {author} {\bibinfo {author} {\bibfnamefont {R.}~\bibnamefont
  {Pastor-Satorras}}\ and\ \bibinfo {author} {\bibfnamefont {A.}~\bibnamefont
  {Vespignani}},\ }\bibfield  {title} {\bibinfo {title} {Epidemic spreading in
  scale-free networks},\ }\href {https://doi.org/10.1103/PhysRevLett.86.3200}
  {\bibfield  {journal} {\bibinfo  {journal} {Physical review letters}\
  }\textbf {\bibinfo {volume} {86}},\ \bibinfo {pages} {3200} (\bibinfo {year}
  {2001})}\BibitemShut {NoStop}%
\bibitem [{\citenamefont {Sahneh}\ \emph {et~al.}(2013)\citenamefont {Sahneh},
  \citenamefont {Scoglio},\ and\ \citenamefont {{Van
  Mieghem}}}]{sahneh2013generalized}%
  \BibitemOpen
  \bibfield  {author} {\bibinfo {author} {\bibfnamefont {F.~D.}\ \bibnamefont
  {Sahneh}}, \bibinfo {author} {\bibfnamefont {C.}~\bibnamefont {Scoglio}},\
  and\ \bibinfo {author} {\bibfnamefont {P.}~\bibnamefont {{Van Mieghem}}},\
  }\bibfield  {title} {\bibinfo {title} {Generalized epidemic mean-field model
  for spreading processes over multilayer complex networks},\ }\href
  {https://doi.org/10.1109/TNET.2013.2239658} {\bibfield  {journal} {\bibinfo
  {journal} {IEEE/ACM Transactions on Networking}\ }\textbf {\bibinfo {volume}
  {21}},\ \bibinfo {pages} {1609–1620} (\bibinfo {year} {2013})}\BibitemShut
  {NoStop}%
\bibitem [{\citenamefont {Pastor-Satorras}\ \emph {et~al.}(2015)\citenamefont
  {Pastor-Satorras}, \citenamefont {Castellano}, \citenamefont {{Van
  Mieghem}},\ and\ \citenamefont {Vespignani}}]{PastorSatorras2015}%
  \BibitemOpen
  \bibfield  {author} {\bibinfo {author} {\bibfnamefont {R.}~\bibnamefont
  {Pastor-Satorras}}, \bibinfo {author} {\bibfnamefont {C.}~\bibnamefont
  {Castellano}}, \bibinfo {author} {\bibfnamefont {P.}~\bibnamefont {{Van
  Mieghem}}},\ and\ \bibinfo {author} {\bibfnamefont {A.}~\bibnamefont
  {Vespignani}},\ }\bibfield  {title} {\bibinfo {title} {Epidemic processes in
  complex networks},\ }\href {https://doi.org/10.1103/RevModPhys.87.925}
  {\bibfield  {journal} {\bibinfo  {journal} {Reviews of modern physics}\
  }\textbf {\bibinfo {volume} {87}},\ \bibinfo {pages} {925} (\bibinfo {year}
  {2015})}\BibitemShut {NoStop}%
\bibitem [{\citenamefont {Merbis}\ and\ \citenamefont
  {de~Domenico}(2023{\natexlab{b}})}]{merbis2023complex}%
  \BibitemOpen
  \bibfield  {author} {\bibinfo {author} {\bibfnamefont {W.}~\bibnamefont
  {Merbis}}\ and\ \bibinfo {author} {\bibfnamefont {M.}~\bibnamefont
  {de~Domenico}},\ }\href@noop {} {\bibinfo {title} {Complex information
  dynamics of epidemic spreading in low-dimensional networks}} (\bibinfo {year}
  {2023}{\natexlab{b}}),\ \Eprint {https://arxiv.org/abs/2305.05429}
  {arXiv:2305.05429 [physics.soc-ph]} \BibitemShut {NoStop}%
\bibitem [{\citenamefont {Poulin}\ \emph {et~al.}(2011)\citenamefont {Poulin},
  \citenamefont {Qarry}, \citenamefont {Somma},\ and\ \citenamefont
  {Verstraete}}]{poulin2011quantum}%
  \BibitemOpen
  \bibfield  {author} {\bibinfo {author} {\bibfnamefont {D.}~\bibnamefont
  {Poulin}}, \bibinfo {author} {\bibfnamefont {A.}~\bibnamefont {Qarry}},
  \bibinfo {author} {\bibfnamefont {R.}~\bibnamefont {Somma}},\ and\ \bibinfo
  {author} {\bibfnamefont {F.}~\bibnamefont {Verstraete}},\ }\bibfield  {title}
  {\bibinfo {title} {Quantum simulation of time-dependent hamiltonians and the
  convenient illusion of hilbert space},\ }\href
  {https://doi.org/10.1103/PhysRevLett.106.170501} {\bibfield  {journal}
  {\bibinfo  {journal} {Physical review letters}\ }\textbf {\bibinfo {volume}
  {106}},\ \bibinfo {pages} {170501} (\bibinfo {year} {2011})}\BibitemShut
  {NoStop}%
\bibitem [{\citenamefont {White}(1992)}]{white1992density}%
  \BibitemOpen
  \bibfield  {author} {\bibinfo {author} {\bibfnamefont {S.~R.}\ \bibnamefont
  {White}},\ }\bibfield  {title} {\bibinfo {title} {Density matrix formulation
  for quantum renormalization groups},\ }\href
  {https://doi.org/10.1103/PhysRevLett.69.2863} {\bibfield  {journal} {\bibinfo
   {journal} {Physical review letters}\ }\textbf {\bibinfo {volume} {69}},\
  \bibinfo {pages} {2863} (\bibinfo {year} {1992})}\BibitemShut {NoStop}%
\bibitem [{\citenamefont {Schollw{\"o}ck}(2011)}]{schollwock2011density}%
  \BibitemOpen
  \bibfield  {author} {\bibinfo {author} {\bibfnamefont {U.}~\bibnamefont
  {Schollw{\"o}ck}},\ }\bibfield  {title} {\bibinfo {title} {The density-matrix
  renormalization group in the age of matrix product states},\ }\href
  {https://doi.org/10.1016/j.aop.2010.09.012} {\bibfield  {journal} {\bibinfo
  {journal} {Annals of physics}\ }\textbf {\bibinfo {volume} {326}},\ \bibinfo
  {pages} {96} (\bibinfo {year} {2011})}\BibitemShut {NoStop}%
\bibitem [{\citenamefont {Orús}(2014)}]{orus2014practical}%
  \BibitemOpen
  \bibfield  {author} {\bibinfo {author} {\bibfnamefont {R.}~\bibnamefont
  {Orús}},\ }\bibfield  {title} {\bibinfo {title} {A practical introduction to
  tensor networks: Matrix product states and projected entangled pair states},\
  }\href {https://doi.org/10.1016/j.aop.2014.06.013} {\bibfield  {journal}
  {\bibinfo  {journal} {Annals of Physics}\ }\textbf {\bibinfo {volume}
  {349}},\ \bibinfo {pages} {117–158} (\bibinfo {year} {2014})}\BibitemShut
  {NoStop}%
\bibitem [{\citenamefont {Sch{\"o}nberg}(1952)}]{schonberg1952application}%
  \BibitemOpen
  \bibfield  {author} {\bibinfo {author} {\bibfnamefont {M.}~\bibnamefont
  {Sch{\"o}nberg}},\ }\bibfield  {title} {\bibinfo {title} {Application of
  second quantization methods to the classical statistical mechanics},\ }\href
  {https://doi.org/10.1007/BF02782925} {\bibfield  {journal} {\bibinfo
  {journal} {Il Nuovo Cimento (1943-1954)}\ }\textbf {\bibinfo {volume} {9}},\
  \bibinfo {pages} {1139} (\bibinfo {year} {1952})}\BibitemShut {NoStop}%
\bibitem [{\citenamefont {Doi}(1976)}]{doi1976second}%
  \BibitemOpen
  \bibfield  {author} {\bibinfo {author} {\bibfnamefont {M.}~\bibnamefont
  {Doi}},\ }\bibfield  {title} {\bibinfo {title} {Second quantization
  representation for classical many-particle system},\ }\href
  {https://doi.org/10.1088/0305-4470/9/9/008} {\bibfield  {journal} {\bibinfo
  {journal} {Journal of Physics A: Mathematical and General}\ }\textbf
  {\bibinfo {volume} {9}},\ \bibinfo {pages} {1465} (\bibinfo {year}
  {1976})}\BibitemShut {NoStop}%
\bibitem [{\citenamefont {Grassberger}\ and\ \citenamefont
  {Scheunert}(1980)}]{grassberger1980fock}%
  \BibitemOpen
  \bibfield  {author} {\bibinfo {author} {\bibfnamefont {P.}~\bibnamefont
  {Grassberger}}\ and\ \bibinfo {author} {\bibfnamefont {M.}~\bibnamefont
  {Scheunert}},\ }\bibfield  {title} {\bibinfo {title} {Fock-space methods for
  identical classical objects},\ }\href
  {https://doi.org/10.1002/prop.19800281004} {\bibfield  {journal} {\bibinfo
  {journal} {Fortschritte der Physik}\ }\textbf {\bibinfo {volume} {28}},\
  \bibinfo {pages} {547} (\bibinfo {year} {1980})}\BibitemShut {NoStop}%
\bibitem [{\citenamefont {Peliti}(1985)}]{peliti1985path}%
  \BibitemOpen
  \bibfield  {author} {\bibinfo {author} {\bibfnamefont {L.}~\bibnamefont
  {Peliti}},\ }\bibfield  {title} {\bibinfo {title} {Path integral approach to
  birth-death processes on a lattice},\ }\href
  {https://doi.org/10.1051/jphys:019850046090146900} {\bibfield  {journal}
  {\bibinfo  {journal} {Journal de Physique}\ }\textbf {\bibinfo {volume}
  {46}},\ \bibinfo {pages} {1469} (\bibinfo {year} {1985})}\BibitemShut
  {NoStop}%
\bibitem [{\citenamefont {Sch{\"u}tz}(1995)}]{schutz1995reaction}%
  \BibitemOpen
  \bibfield  {author} {\bibinfo {author} {\bibfnamefont {G.~M.}\ \bibnamefont
  {Sch{\"u}tz}},\ }\bibfield  {title} {\bibinfo {title} {Reaction-diffusion
  processes of hard-core particles},\ }\href
  {https://doi.org/10.1007/BF02179389} {\bibfield  {journal} {\bibinfo
  {journal} {Journal of statistical physics}\ }\textbf {\bibinfo {volume}
  {79}},\ \bibinfo {pages} {243} (\bibinfo {year} {1995})}\BibitemShut
  {NoStop}%
\bibitem [{\citenamefont {Hinrichsen}(2000)}]{hinrichsen2000non}%
  \BibitemOpen
  \bibfield  {author} {\bibinfo {author} {\bibfnamefont {H.}~\bibnamefont
  {Hinrichsen}},\ }\bibfield  {title} {\bibinfo {title} {Non-equilibrium
  critical phenomena and phase transitions into absorbing states},\ }\href@noop
  {} {\bibfield  {journal} {\bibinfo  {journal} {Advances in physics}\ }\textbf
  {\bibinfo {volume} {49}},\ \bibinfo {pages} {815} (\bibinfo {year}
  {2000})}\BibitemShut {NoStop}%
\bibitem [{\citenamefont {Schütz}(2001)}]{SCHUTZ20011}%
  \BibitemOpen
  \bibfield  {author} {\bibinfo {author} {\bibfnamefont {G.}~\bibnamefont
  {Schütz}},\ }\bibfield  {title} {\bibinfo {title} {Exactly solvable models
  for many-body systems far from equilibrium}\ }(\bibinfo  {publisher}
  {Academic Press},\ \bibinfo {year} {2001})\ pp.\ \bibinfo {pages}
  {1--251}\BibitemShut {NoStop}%
\bibitem [{\citenamefont {Merbis}\ and\ \citenamefont
  {Lodato}(2022)}]{merbis2022logistic}%
  \BibitemOpen
  \bibfield  {author} {\bibinfo {author} {\bibfnamefont {W.}~\bibnamefont
  {Merbis}}\ and\ \bibinfo {author} {\bibfnamefont {I.}~\bibnamefont
  {Lodato}},\ }\bibfield  {title} {\bibinfo {title} {Logistic growth on
  networks: Exact solutions for the susceptible-infected model},\ }\href
  {https://doi.org/10.1103/PhysRevE.105.044303} {\bibfield  {journal} {\bibinfo
   {journal} {Physical Review E}\ }\textbf {\bibinfo {volume} {105}},\ \bibinfo
  {pages} {044303} (\bibinfo {year} {2022})}\BibitemShut {NoStop}%
\bibitem [{\citenamefont {Derrida}\ and\ \citenamefont
  {Lebowitz}(1998)}]{derrida1998exact}%
  \BibitemOpen
  \bibfield  {author} {\bibinfo {author} {\bibfnamefont {B.}~\bibnamefont
  {Derrida}}\ and\ \bibinfo {author} {\bibfnamefont {J.~L.}\ \bibnamefont
  {Lebowitz}},\ }\bibfield  {title} {\bibinfo {title} {Exact large deviation
  function in the asymmetric exclusion process},\ }\href
  {https://doi.org/10.1103/PhysRevLett.80.209} {\bibfield  {journal} {\bibinfo
  {journal} {Physical review letters}\ }\textbf {\bibinfo {volume} {80}},\
  \bibinfo {pages} {209} (\bibinfo {year} {1998})}\BibitemShut {NoStop}%
\bibitem [{\citenamefont {Johnson}\ \emph {et~al.}(2010)\citenamefont
  {Johnson}, \citenamefont {Clark},\ and\ \citenamefont
  {Jaksch}}]{Johnson2010}%
  \BibitemOpen
  \bibfield  {author} {\bibinfo {author} {\bibfnamefont {T.}~\bibnamefont
  {Johnson}}, \bibinfo {author} {\bibfnamefont {S.}~\bibnamefont {Clark}},\
  and\ \bibinfo {author} {\bibfnamefont {D.}~\bibnamefont {Jaksch}},\
  }\bibfield  {title} {\bibinfo {title} {Dynamical simulations of classical
  stochastic systems using matrix product states},\ }\href
  {https://doi.org/10.1103/PhysRevE.82.036702} {\bibfield  {journal} {\bibinfo
  {journal} {Physical Review E}\ }\textbf {\bibinfo {volume} {82}},\ \bibinfo
  {pages} {036702} (\bibinfo {year} {2010})}\BibitemShut {NoStop}%
\bibitem [{\citenamefont {de~Gier}\ and\ \citenamefont
  {Essler}(2011)}]{de2011large}%
  \BibitemOpen
  \bibfield  {author} {\bibinfo {author} {\bibfnamefont {J.}~\bibnamefont
  {de~Gier}}\ and\ \bibinfo {author} {\bibfnamefont {F.~H.}\ \bibnamefont
  {Essler}},\ }\bibfield  {title} {\bibinfo {title} {Large deviation function
  for the current in the open asymmetric simple exclusion process},\ }\href
  {https://doi.org/10.1103/PhysRevLett.107.010602} {\bibfield  {journal}
  {\bibinfo  {journal} {Physical review letters}\ }\textbf {\bibinfo {volume}
  {107}},\ \bibinfo {pages} {010602} (\bibinfo {year} {2011})}\BibitemShut
  {NoStop}%
\bibitem [{\citenamefont {Lazarescu}\ and\ \citenamefont
  {Mallick}(2011)}]{lazarescu2011exact}%
  \BibitemOpen
  \bibfield  {author} {\bibinfo {author} {\bibfnamefont {A.}~\bibnamefont
  {Lazarescu}}\ and\ \bibinfo {author} {\bibfnamefont {K.}~\bibnamefont
  {Mallick}},\ }\bibfield  {title} {\bibinfo {title} {An exact formula for the
  statistics of the current in the tasep with open boundaries},\ }\href
  {https://doi.org/10.1088/1751-8113/44/31/315001} {\bibfield  {journal}
  {\bibinfo  {journal} {Journal of Physics A: Mathematical and Theoretical}\
  }\textbf {\bibinfo {volume} {44}},\ \bibinfo {pages} {315001} (\bibinfo
  {year} {2011})}\BibitemShut {NoStop}%
\bibitem [{\citenamefont {Gorissen}\ \emph {et~al.}(2012)\citenamefont
  {Gorissen}, \citenamefont {Lazarescu}, \citenamefont {Mallick},\ and\
  \citenamefont {Vanderzande}}]{gorissen2012exact}%
  \BibitemOpen
  \bibfield  {author} {\bibinfo {author} {\bibfnamefont {M.}~\bibnamefont
  {Gorissen}}, \bibinfo {author} {\bibfnamefont {A.}~\bibnamefont {Lazarescu}},
  \bibinfo {author} {\bibfnamefont {K.}~\bibnamefont {Mallick}},\ and\ \bibinfo
  {author} {\bibfnamefont {C.}~\bibnamefont {Vanderzande}},\ }\bibfield
  {title} {\bibinfo {title} {Exact current statistics of the asymmetric simple
  exclusion process with open boundaries},\ }\href
  {https://doi.org/10.1103/PhysRevLett.109.170601} {\bibfield  {journal}
  {\bibinfo  {journal} {Physical review letters}\ }\textbf {\bibinfo {volume}
  {109}},\ \bibinfo {pages} {170601} (\bibinfo {year} {2012})}\BibitemShut
  {NoStop}%
\bibitem [{\citenamefont {Nishino}(1995)}]{nishino1995density}%
  \BibitemOpen
  \bibfield  {author} {\bibinfo {author} {\bibfnamefont {T.}~\bibnamefont
  {Nishino}},\ }\bibfield  {title} {\bibinfo {title} {Density matrix
  renormalization group method for 2d classical models},\ }\href
  {https://journals.jps.jp/doi/pdf/10.1143/JPSJ.64.3598} {\bibfield  {journal}
  {\bibinfo  {journal} {Journal of the Physical Society of Japan}\ }\textbf
  {\bibinfo {volume} {64}},\ \bibinfo {pages} {3598} (\bibinfo {year}
  {1995})}\BibitemShut {NoStop}%
\bibitem [{\citenamefont {Carlon}\ \emph {et~al.}(1999)\citenamefont {Carlon},
  \citenamefont {Henkel},\ and\ \citenamefont
  {Schollw{\"o}ck}}]{carlon1999density}%
  \BibitemOpen
  \bibfield  {author} {\bibinfo {author} {\bibfnamefont {E.}~\bibnamefont
  {Carlon}}, \bibinfo {author} {\bibfnamefont {M.}~\bibnamefont {Henkel}},\
  and\ \bibinfo {author} {\bibfnamefont {U.}~\bibnamefont {Schollw{\"o}ck}},\
  }\bibfield  {title} {\bibinfo {title} {Density matrix renormalization group
  and reaction-diffusion processes},\ }\href
  {https://doi.org/10.1007/s100510050983} {\bibfield  {journal} {\bibinfo
  {journal} {The European Physical Journal B-Condensed Matter and Complex
  Systems}\ }\textbf {\bibinfo {volume} {12}},\ \bibinfo {pages} {99} (\bibinfo
  {year} {1999})}\BibitemShut {NoStop}%
\bibitem [{\citenamefont {Gorissen}\ \emph {et~al.}(2009)\citenamefont
  {Gorissen}, \citenamefont {Hooyberghs},\ and\ \citenamefont
  {Vanderzande}}]{gorissen2009density}%
  \BibitemOpen
  \bibfield  {author} {\bibinfo {author} {\bibfnamefont {M.}~\bibnamefont
  {Gorissen}}, \bibinfo {author} {\bibfnamefont {J.}~\bibnamefont
  {Hooyberghs}},\ and\ \bibinfo {author} {\bibfnamefont {C.}~\bibnamefont
  {Vanderzande}},\ }\bibfield  {title} {\bibinfo {title} {Density-matrix
  renormalization-group study of current and activity fluctuations near
  nonequilibrium phase transitions},\ }\href
  {http://dx.doi.org/10.1103/PhysRevE.79.020101} {\bibfield  {journal}
  {\bibinfo  {journal} {Physical Review E}\ }\textbf {\bibinfo {volume} {79}},\
  \bibinfo {pages} {020101} (\bibinfo {year} {2009})}\BibitemShut {NoStop}%
\bibitem [{\citenamefont {Hooyberghs}\ and\ \citenamefont
  {Vanderzande}(2010)}]{hooyberghs2010thermodynamics}%
  \BibitemOpen
  \bibfield  {author} {\bibinfo {author} {\bibfnamefont {J.}~\bibnamefont
  {Hooyberghs}}\ and\ \bibinfo {author} {\bibfnamefont {C.}~\bibnamefont
  {Vanderzande}},\ }\bibfield  {title} {\bibinfo {title} {Thermodynamics of
  histories for the one-dimensional contact process},\ }\href
  {https://doi.org/10.1088/1742-5468/2010/02/P02017} {\bibfield  {journal}
  {\bibinfo  {journal} {Journal of Statistical Mechanics: Theory and
  Experiment}\ }\textbf {\bibinfo {volume} {2010}},\ \bibinfo {pages} {P02017}
  (\bibinfo {year} {2010})}\BibitemShut {NoStop}%
\bibitem [{\citenamefont {Helms}\ and\ \citenamefont
  {Chan}(2020)}]{helms2020dynamical}%
  \BibitemOpen
  \bibfield  {author} {\bibinfo {author} {\bibfnamefont {P.}~\bibnamefont
  {Helms}}\ and\ \bibinfo {author} {\bibfnamefont {G.~K.-L.}\ \bibnamefont
  {Chan}},\ }\bibfield  {title} {\bibinfo {title} {Dynamical phase transitions
  in a 2d classical nonequilibrium model via 2d tensor networks},\ }\href
  {https://doi.org/10.1103/PhysRevLett.125.140601} {\bibfield  {journal}
  {\bibinfo  {journal} {Physical Review Letters}\ }\textbf {\bibinfo {volume}
  {125}},\ \bibinfo {pages} {140601} (\bibinfo {year} {2020})}\BibitemShut
  {NoStop}%
\bibitem [{\citenamefont {Causer}\ \emph
  {et~al.}(2022{\natexlab{a}})\citenamefont {Causer}, \citenamefont
  {Ba{\~n}uls},\ and\ \citenamefont {Garrahan}}]{causer2022optimal}%
  \BibitemOpen
  \bibfield  {author} {\bibinfo {author} {\bibfnamefont {L.}~\bibnamefont
  {Causer}}, \bibinfo {author} {\bibfnamefont {M.~C.}\ \bibnamefont
  {Ba{\~n}uls}},\ and\ \bibinfo {author} {\bibfnamefont {J.~P.}\ \bibnamefont
  {Garrahan}},\ }\bibfield  {title} {\bibinfo {title} {Optimal sampling of
  dynamical large deviations in two dimensions via tensor networks},\ }\href
  {https://arxiv.org/abs/2209.11788} {\bibfield  {journal} {\bibinfo  {journal}
  {arXiv preprint arXiv:2209.11788}\ } (\bibinfo {year}
  {2022}{\natexlab{a}})}\BibitemShut {NoStop}%
\bibitem [{\citenamefont {Helms}\ \emph {et~al.}(2019)\citenamefont {Helms},
  \citenamefont {Ray},\ and\ \citenamefont {Chan}}]{helms2019dynamical}%
  \BibitemOpen
  \bibfield  {author} {\bibinfo {author} {\bibfnamefont {P.}~\bibnamefont
  {Helms}}, \bibinfo {author} {\bibfnamefont {U.}~\bibnamefont {Ray}},\ and\
  \bibinfo {author} {\bibfnamefont {G.~K.-L.}\ \bibnamefont {Chan}},\
  }\bibfield  {title} {\bibinfo {title} {Dynamical phase behavior of the
  single-and multi-lane asymmetric simple exclusion process via matrix product
  states},\ }\href {https://doi.org/10.1103/PhysRevE.100.022101} {\bibfield
  {journal} {\bibinfo  {journal} {Physical Review E}\ }\textbf {\bibinfo
  {volume} {100}},\ \bibinfo {pages} {022101} (\bibinfo {year}
  {2019})}\BibitemShut {NoStop}%
\bibitem [{\citenamefont {Ba{\~n}uls}\ and\ \citenamefont
  {Garrahan}(2019)}]{banuls2019using}%
  \BibitemOpen
  \bibfield  {author} {\bibinfo {author} {\bibfnamefont {M.~C.}\ \bibnamefont
  {Ba{\~n}uls}}\ and\ \bibinfo {author} {\bibfnamefont {J.~P.}\ \bibnamefont
  {Garrahan}},\ }\bibfield  {title} {\bibinfo {title} {Using matrix product
  states to study the dynamical large deviations of kinetically constrained
  models},\ }\href {https://doi.org/10.1103/PhysRevLett.123.200601} {\bibfield
  {journal} {\bibinfo  {journal} {Physical review letters}\ }\textbf {\bibinfo
  {volume} {123}},\ \bibinfo {pages} {200601} (\bibinfo {year}
  {2019})}\BibitemShut {NoStop}%
\bibitem [{\citenamefont {Causer}\ \emph
  {et~al.}(2022{\natexlab{b}})\citenamefont {Causer}, \citenamefont {Banuls},\
  and\ \citenamefont {Garrahan}}]{causer2022finite}%
  \BibitemOpen
  \bibfield  {author} {\bibinfo {author} {\bibfnamefont {L.}~\bibnamefont
  {Causer}}, \bibinfo {author} {\bibfnamefont {M.~C.}\ \bibnamefont {Banuls}},\
  and\ \bibinfo {author} {\bibfnamefont {J.~P.}\ \bibnamefont {Garrahan}},\
  }\bibfield  {title} {\bibinfo {title} {Finite time large deviations via
  matrix product states},\ }\href
  {https://doi.org/10.1103/PhysRevLett.128.090605} {\bibfield  {journal}
  {\bibinfo  {journal} {Physical Review Letters}\ }\textbf {\bibinfo {volume}
  {128}},\ \bibinfo {pages} {090605} (\bibinfo {year}
  {2022}{\natexlab{b}})}\BibitemShut {NoStop}%
\bibitem [{\citenamefont {Strand}\ \emph {et~al.}(2022)\citenamefont {Strand},
  \citenamefont {Vroylandt},\ and\ \citenamefont {Gingrich}}]{strand2022using}%
  \BibitemOpen
  \bibfield  {author} {\bibinfo {author} {\bibfnamefont {N.~E.}\ \bibnamefont
  {Strand}}, \bibinfo {author} {\bibfnamefont {H.}~\bibnamefont {Vroylandt}},\
  and\ \bibinfo {author} {\bibfnamefont {T.~R.}\ \bibnamefont {Gingrich}},\
  }\bibfield  {title} {\bibinfo {title} {Using tensor network states for
  multi-particle brownian ratchets},\ }\href
  {https://arxiv.org/pdf/2201.03531.pdf} {\bibfield  {journal} {\bibinfo
  {journal} {arXiv preprint arXiv:2201.03531}\ } (\bibinfo {year}
  {2022})}\BibitemShut {NoStop}%
\bibitem [{\citenamefont {Garrahan}\ and\ \citenamefont
  {Pollmann}(2022)}]{garrahan2022topological}%
  \BibitemOpen
  \bibfield  {author} {\bibinfo {author} {\bibfnamefont {J.~P.}\ \bibnamefont
  {Garrahan}}\ and\ \bibinfo {author} {\bibfnamefont {F.}~\bibnamefont
  {Pollmann}},\ }\bibfield  {title} {\bibinfo {title} {Topological phases in
  the dynamics of the simple exclusion process},\ }\href
  {https://arxiv.org/pdf/2203.08200.pdf} {\bibfield  {journal} {\bibinfo
  {journal} {arXiv preprint arXiv:2203.08200}\ } (\bibinfo {year}
  {2022})}\BibitemShut {NoStop}%
\bibitem [{\citenamefont {Johnson}\ \emph {et~al.}(2015)\citenamefont
  {Johnson}, \citenamefont {Elliott}, \citenamefont {Clark},\ and\
  \citenamefont {Jaksch}}]{Johnson2014}%
  \BibitemOpen
  \bibfield  {author} {\bibinfo {author} {\bibfnamefont {T.~H.}\ \bibnamefont
  {Johnson}}, \bibinfo {author} {\bibfnamefont {T.}~\bibnamefont {Elliott}},
  \bibinfo {author} {\bibfnamefont {S.~R.}\ \bibnamefont {Clark}},\ and\
  \bibinfo {author} {\bibfnamefont {D.}~\bibnamefont {Jaksch}},\ }\bibfield
  {title} {\bibinfo {title} {Capturing exponential variance using polynomial
  resources: applying tensor networks to nonequilibrium stochastic processes},\
  }\href {https://doi.org/10.1103/PhysRevLett.114.090602} {\bibfield  {journal}
  {\bibinfo  {journal} {Physical review letters}\ }\textbf {\bibinfo {volume}
  {114}},\ \bibinfo {pages} {090602} (\bibinfo {year} {2015})}\BibitemShut
  {NoStop}%
\bibitem [{\citenamefont {Causer}\ \emph {et~al.}(2021)\citenamefont {Causer},
  \citenamefont {Ba{\~n}uls},\ and\ \citenamefont
  {Garrahan}}]{causer2021optimal}%
  \BibitemOpen
  \bibfield  {author} {\bibinfo {author} {\bibfnamefont {L.}~\bibnamefont
  {Causer}}, \bibinfo {author} {\bibfnamefont {M.~C.}\ \bibnamefont
  {Ba{\~n}uls}},\ and\ \bibinfo {author} {\bibfnamefont {J.~P.}\ \bibnamefont
  {Garrahan}},\ }\bibfield  {title} {\bibinfo {title} {Optimal sampling of
  dynamical large deviations via matrix product states},\ }\href
  {https://doi.org/10.1103/PhysRevE.103.062144} {\bibfield  {journal} {\bibinfo
   {journal} {Physical Review E}\ }\textbf {\bibinfo {volume} {103}},\ \bibinfo
  {pages} {062144} (\bibinfo {year} {2021})}\BibitemShut {NoStop}%
\bibitem [{\citenamefont {Harris}(1974)}]{harris1974contact}%
  \BibitemOpen
  \bibfield  {author} {\bibinfo {author} {\bibfnamefont {T.~E.}\ \bibnamefont
  {Harris}},\ }\bibfield  {title} {\bibinfo {title} {Contact interactions on a
  lattice},\ }\href {https://doi.org/10.1214/aop/1176996493} {\bibfield
  {journal} {\bibinfo  {journal} {The Annals of Probability}\ }\textbf
  {\bibinfo {volume} {2}},\ \bibinfo {pages} {969} (\bibinfo {year}
  {1974})}\BibitemShut {NoStop}%
\bibitem [{\citenamefont {Grassberger}(1983)}]{grassberger1983critical}%
  \BibitemOpen
  \bibfield  {author} {\bibinfo {author} {\bibfnamefont {P.}~\bibnamefont
  {Grassberger}},\ }\bibfield  {title} {\bibinfo {title} {On the critical
  behavior of the general epidemic process and dynamical percolation},\ }\href
  {https://doi.org/10.1016/0025-5564(82)90036-0} {\bibfield  {journal}
  {\bibinfo  {journal} {Mathematical Biosciences}\ }\textbf {\bibinfo {volume}
  {63}},\ \bibinfo {pages} {157–172} (\bibinfo {year} {1983})}\BibitemShut
  {NoStop}%
\bibitem [{\citenamefont {Jensen}\ and\ \citenamefont
  {Dickman}(1993)}]{jensen1993time}%
  \BibitemOpen
  \bibfield  {author} {\bibinfo {author} {\bibfnamefont {I.}~\bibnamefont
  {Jensen}}\ and\ \bibinfo {author} {\bibfnamefont {R.}~\bibnamefont
  {Dickman}},\ }\bibfield  {title} {\bibinfo {title} {Time-dependent
  perturbation theory for nonequilibrium lattice models},\ }\href
  {https://doi.org/10.1007/BF01048090} {\bibfield  {journal} {\bibinfo
  {journal} {Journal of statistical physics}\ }\textbf {\bibinfo {volume}
  {71}},\ \bibinfo {pages} {89} (\bibinfo {year} {1993})}\BibitemShut {NoStop}%
\bibitem [{\citenamefont {Hooyberghs}\ and\ \citenamefont
  {Vanderzande}(2001)}]{hooyberghs2001one}%
  \BibitemOpen
  \bibfield  {author} {\bibinfo {author} {\bibfnamefont {J.}~\bibnamefont
  {Hooyberghs}}\ and\ \bibinfo {author} {\bibfnamefont {C.}~\bibnamefont
  {Vanderzande}},\ }\bibfield  {title} {\bibinfo {title} {One-dimensional
  contact process: Duality and renormalization},\ }\href
  {https://doi.org/10.1103/PhysRevE.63.041109} {\bibfield  {journal} {\bibinfo
  {journal} {Physical Review E}\ }\textbf {\bibinfo {volume} {63}},\ \bibinfo
  {pages} {041109} (\bibinfo {year} {2001})}\BibitemShut {NoStop}%
\bibitem [{\citenamefont {Marro}\ and\ \citenamefont
  {Dickman}(2005)}]{marro2005nonequilibrium}%
  \BibitemOpen
  \bibfield  {author} {\bibinfo {author} {\bibfnamefont {J.}~\bibnamefont
  {Marro}}\ and\ \bibinfo {author} {\bibfnamefont {R.}~\bibnamefont
  {Dickman}},\ }\href@noop {} {\emph {\bibinfo {title} {Nonequilibrium Phase
  Transitions in Lattice Models}}}\ (\bibinfo  {publisher} {Cambridge
  University Press},\ \bibinfo {year} {2005})\BibitemShut {NoStop}%
\bibitem [{\citenamefont {Ramola}\ \emph {et~al.}(2014)\citenamefont {Ramola},
  \citenamefont {Majumdar},\ and\ \citenamefont
  {Schehr}}]{ramola2014universal}%
  \BibitemOpen
  \bibfield  {author} {\bibinfo {author} {\bibfnamefont {K.}~\bibnamefont
  {Ramola}}, \bibinfo {author} {\bibfnamefont {S.~N.}\ \bibnamefont
  {Majumdar}},\ and\ \bibinfo {author} {\bibfnamefont {G.}~\bibnamefont
  {Schehr}},\ }\bibfield  {title} {\bibinfo {title} {Universal order and gap
  statistics of critical branching brownian motion},\ }\href@noop {} {\bibfield
   {journal} {\bibinfo  {journal} {Physical Review Letters}\ }\textbf {\bibinfo
  {volume} {112}},\ \bibinfo {pages} {210602} (\bibinfo {year}
  {2014})}\BibitemShut {NoStop}%
\bibitem [{\citenamefont {Lecomte}\ \emph {et~al.}(2007)\citenamefont
  {Lecomte}, \citenamefont {Appert-Rolland},\ and\ \citenamefont
  {Van~Wijland}}]{lecomte2007thermodynamic}%
  \BibitemOpen
  \bibfield  {author} {\bibinfo {author} {\bibfnamefont {V.}~\bibnamefont
  {Lecomte}}, \bibinfo {author} {\bibfnamefont {C.}~\bibnamefont
  {Appert-Rolland}},\ and\ \bibinfo {author} {\bibfnamefont {F.}~\bibnamefont
  {Van~Wijland}},\ }\bibfield  {title} {\bibinfo {title} {Thermodynamic
  formalism for systems with markov dynamics},\ }\href
  {https://doi.org/10.1007/s10955-006-9254-0} {\bibfield  {journal} {\bibinfo
  {journal} {Journal of statistical physics}\ }\textbf {\bibinfo {volume}
  {127}},\ \bibinfo {pages} {51} (\bibinfo {year} {2007})}\BibitemShut
  {NoStop}%
\bibitem [{\citenamefont {Garrahan}\ \emph {et~al.}(2009)\citenamefont
  {Garrahan}, \citenamefont {Jack}, \citenamefont {Lecomte}, \citenamefont
  {Pitard}, \citenamefont {van Duijvendijk},\ and\ \citenamefont {van
  Wijland}}]{garrahan2009first}%
  \BibitemOpen
  \bibfield  {author} {\bibinfo {author} {\bibfnamefont {J.~P.}\ \bibnamefont
  {Garrahan}}, \bibinfo {author} {\bibfnamefont {R.~L.}\ \bibnamefont {Jack}},
  \bibinfo {author} {\bibfnamefont {V.}~\bibnamefont {Lecomte}}, \bibinfo
  {author} {\bibfnamefont {E.}~\bibnamefont {Pitard}}, \bibinfo {author}
  {\bibfnamefont {K.}~\bibnamefont {van Duijvendijk}},\ and\ \bibinfo {author}
  {\bibfnamefont {F.}~\bibnamefont {van Wijland}},\ }\bibfield  {title}
  {\bibinfo {title} {First-order dynamical phase transition in models of
  glasses: an approach based on ensembles of histories},\ }\href
  {https://doi.org/10.1088/1751-8113/42/7/075007} {\bibfield  {journal}
  {\bibinfo  {journal} {Journal of Physics A: Mathematical and Theoretical}\
  }\textbf {\bibinfo {volume} {42}},\ \bibinfo {pages} {075007} (\bibinfo
  {year} {2009})}\BibitemShut {NoStop}%
\bibitem [{\citenamefont {Chetrite}\ and\ \citenamefont
  {Touchette}(2015)}]{chetrite2015nonequilibrium}%
  \BibitemOpen
  \bibfield  {author} {\bibinfo {author} {\bibfnamefont {R.}~\bibnamefont
  {Chetrite}}\ and\ \bibinfo {author} {\bibfnamefont {H.}~\bibnamefont
  {Touchette}},\ }\bibfield  {title} {\bibinfo {title} {Nonequilibrium markov
  processes conditioned on large deviations},\ }in\ \href
  {https://doi.org/10.1007/s00023-014-0375-8} {\emph {\bibinfo {booktitle}
  {Annales Henri Poincar{\'e}}}},\ Vol.~\bibinfo {volume} {16}\ (\bibinfo
  {organization} {Springer},\ \bibinfo {year} {2015})\ pp.\ \bibinfo {pages}
  {2005--2057}\BibitemShut {NoStop}%
\bibitem [{\citenamefont {Touchette}(2009)}]{touchette2009large}%
  \BibitemOpen
  \bibfield  {author} {\bibinfo {author} {\bibfnamefont {H.}~\bibnamefont
  {Touchette}},\ }\bibfield  {title} {\bibinfo {title} {The large deviation
  approach to statistical mechanics},\ }\href
  {https://doi.org/10.1016/j.physrep.2009.05.002} {\bibfield  {journal}
  {\bibinfo  {journal} {Physics Reports}\ }\textbf {\bibinfo {volume} {478}},\
  \bibinfo {pages} {1} (\bibinfo {year} {2009})}\BibitemShut {NoStop}%
\bibitem [{\citenamefont {Wolf}\ \emph {et~al.}(2008)\citenamefont {Wolf},
  \citenamefont {Verstraete}, \citenamefont {Hastings},\ and\ \citenamefont
  {Cirac}}]{wolf2008area}%
  \BibitemOpen
  \bibfield  {author} {\bibinfo {author} {\bibfnamefont {M.~M.}\ \bibnamefont
  {Wolf}}, \bibinfo {author} {\bibfnamefont {F.}~\bibnamefont {Verstraete}},
  \bibinfo {author} {\bibfnamefont {M.~B.}\ \bibnamefont {Hastings}},\ and\
  \bibinfo {author} {\bibfnamefont {J.~I.}\ \bibnamefont {Cirac}},\ }\bibfield
  {title} {\bibinfo {title} {Area laws in quantum systems: mutual information
  and correlations},\ }\href
  {https://journals.aps.org/prl/abstract/10.1103/PhysRevLett.100.070502}
  {\bibfield  {journal} {\bibinfo  {journal} {Physical review letters}\
  }\textbf {\bibinfo {volume} {100}},\ \bibinfo {pages} {070502} (\bibinfo
  {year} {2008})}\BibitemShut {NoStop}%
\bibitem [{\citenamefont {Schuch}\ \emph {et~al.}(2008)\citenamefont {Schuch},
  \citenamefont {Wolf}, \citenamefont {Verstraete},\ and\ \citenamefont
  {Cirac}}]{schuch2008entropy}%
  \BibitemOpen
  \bibfield  {author} {\bibinfo {author} {\bibfnamefont {N.}~\bibnamefont
  {Schuch}}, \bibinfo {author} {\bibfnamefont {M.~M.}\ \bibnamefont {Wolf}},
  \bibinfo {author} {\bibfnamefont {F.}~\bibnamefont {Verstraete}},\ and\
  \bibinfo {author} {\bibfnamefont {J.~I.}\ \bibnamefont {Cirac}},\ }\bibfield
  {title} {\bibinfo {title} {Entropy scaling and simulability by matrix product
  states},\ }\href {https://doi.org/10.1103/PhysRevLett.100.070502} {\bibfield
  {journal} {\bibinfo  {journal} {Physical review letters}\ }\textbf {\bibinfo
  {volume} {100}},\ \bibinfo {pages} {030504} (\bibinfo {year}
  {2008})}\BibitemShut {NoStop}%
\bibitem [{\citenamefont {Scalet}\ \emph {et~al.}(2021)\citenamefont {Scalet},
  \citenamefont {Alhambra}, \citenamefont {Styliaris},\ and\ \citenamefont
  {Cirac}}]{scalet2021computable}%
  \BibitemOpen
  \bibfield  {author} {\bibinfo {author} {\bibfnamefont {S.~O.}\ \bibnamefont
  {Scalet}}, \bibinfo {author} {\bibfnamefont {{\'A}.~M.}\ \bibnamefont
  {Alhambra}}, \bibinfo {author} {\bibfnamefont {G.}~\bibnamefont
  {Styliaris}},\ and\ \bibinfo {author} {\bibfnamefont {J.~I.}\ \bibnamefont
  {Cirac}},\ }\bibfield  {title} {\bibinfo {title} {Computable r{\'e}nyi mutual
  information: Area laws and correlations},\ }\href
  {https://doi.org/10.22331/q-2021-09-14-541} {\bibfield  {journal} {\bibinfo
  {journal} {Quantum}\ }\textbf {\bibinfo {volume} {5}},\ \bibinfo {pages}
  {541} (\bibinfo {year} {2021})}\BibitemShut {NoStop}%
\bibitem [{\citenamefont {Iaconis}\ \emph {et~al.}(2013)\citenamefont
  {Iaconis}, \citenamefont {Inglis}, \citenamefont {Kallin},\ and\
  \citenamefont {Melko}}]{iaconis2013detecting}%
  \BibitemOpen
  \bibfield  {author} {\bibinfo {author} {\bibfnamefont {J.}~\bibnamefont
  {Iaconis}}, \bibinfo {author} {\bibfnamefont {S.}~\bibnamefont {Inglis}},
  \bibinfo {author} {\bibfnamefont {A.~B.}\ \bibnamefont {Kallin}},\ and\
  \bibinfo {author} {\bibfnamefont {R.~G.}\ \bibnamefont {Melko}},\ }\bibfield
  {title} {\bibinfo {title} {Detecting classical phase transitions with renyi
  mutual information},\ }\href {https://doi.org/10.1103/PhysRevB.87.195134}
  {\bibfield  {journal} {\bibinfo  {journal} {Physical Review B}\ }\textbf
  {\bibinfo {volume} {87}},\ \bibinfo {pages} {195134} (\bibinfo {year}
  {2013})}\BibitemShut {NoStop}%
\bibitem [{\citenamefont {Merbis}(2021)}]{merbis2021exact}%
  \BibitemOpen
  \bibfield  {author} {\bibinfo {author} {\bibfnamefont {W.}~\bibnamefont
  {Merbis}},\ }\bibfield  {title} {\bibinfo {title} {Exact epidemic models from
  a tensor product formulation},\ }\href {https://arxiv.org/abs/2102.11708}
  {\bibfield  {journal} {\bibinfo  {journal} {arXiv:2102.11708}\ } (\bibinfo
  {year} {2021})}\BibitemShut {NoStop}%
\bibitem [{\citenamefont {de~Mendon{\c{c}}a}(1999)}]{de1999precise}%
  \BibitemOpen
  \bibfield  {author} {\bibinfo {author} {\bibfnamefont {J.~R.~G.}\
  \bibnamefont {de~Mendon{\c{c}}a}},\ }\bibfield  {title} {\bibinfo {title}
  {Precise critical exponents for the basic contact process},\ }\href
  {https://doi.org/10.1088/0305-4470/32/44/102} {\bibfield  {journal} {\bibinfo
   {journal} {Journal of Physics A: Mathematical and General}\ }\textbf
  {\bibinfo {volume} {32}},\ \bibinfo {pages} {L467} (\bibinfo {year}
  {1999})}\BibitemShut {NoStop}%
\bibitem [{\citenamefont {Eisert}\ \emph {et~al.}(2010)\citenamefont {Eisert},
  \citenamefont {Cramer},\ and\ \citenamefont {Plenio}}]{eisert2010colloquium}%
  \BibitemOpen
  \bibfield  {author} {\bibinfo {author} {\bibfnamefont {J.}~\bibnamefont
  {Eisert}}, \bibinfo {author} {\bibfnamefont {M.}~\bibnamefont {Cramer}},\
  and\ \bibinfo {author} {\bibfnamefont {M.~B.}\ \bibnamefont {Plenio}},\
  }\bibfield  {title} {\bibinfo {title} {Colloquium: Area laws for the
  entanglement entropy},\ }\href@noop {} {\bibfield  {journal} {\bibinfo
  {journal} {Reviews of modern physics}\ }\textbf {\bibinfo {volume} {82}},\
  \bibinfo {pages} {277} (\bibinfo {year} {2010})}\BibitemShut {NoStop}%
\bibitem [{\citenamefont {Garrahan}\ \emph {et~al.}(2007)\citenamefont
  {Garrahan}, \citenamefont {Jack}, \citenamefont {Lecomte}, \citenamefont
  {Pitard}, \citenamefont {van Duijvendijk},\ and\ \citenamefont {van
  Wijland}}]{garrahan2007dynamical}%
  \BibitemOpen
  \bibfield  {author} {\bibinfo {author} {\bibfnamefont {J.~P.}\ \bibnamefont
  {Garrahan}}, \bibinfo {author} {\bibfnamefont {R.~L.}\ \bibnamefont {Jack}},
  \bibinfo {author} {\bibfnamefont {V.}~\bibnamefont {Lecomte}}, \bibinfo
  {author} {\bibfnamefont {E.}~\bibnamefont {Pitard}}, \bibinfo {author}
  {\bibfnamefont {K.}~\bibnamefont {van Duijvendijk}},\ and\ \bibinfo {author}
  {\bibfnamefont {F.}~\bibnamefont {van Wijland}},\ }\bibfield  {title}
  {\bibinfo {title} {Dynamical first-order phase transition in kinetically
  constrained models of glasses},\ }\href
  {https://doi.org/10.1103/PhysRevLett.98.195702} {\bibfield  {journal}
  {\bibinfo  {journal} {Physical review letters}\ }\textbf {\bibinfo {volume}
  {98}},\ \bibinfo {pages} {195702} (\bibinfo {year} {2007})}\BibitemShut
  {NoStop}%
\bibitem [{\citenamefont {Verstraete}\ \emph {et~al.}(2008)\citenamefont
  {Verstraete}, \citenamefont {Murg},\ and\ \citenamefont
  {Cirac}}]{verstraete2008matrix}%
  \BibitemOpen
  \bibfield  {author} {\bibinfo {author} {\bibfnamefont {F.}~\bibnamefont
  {Verstraete}}, \bibinfo {author} {\bibfnamefont {V.}~\bibnamefont {Murg}},\
  and\ \bibinfo {author} {\bibfnamefont {J.~I.}\ \bibnamefont {Cirac}},\
  }\bibfield  {title} {\bibinfo {title} {Matrix product states, projected
  entangled pair states, and variational renormalization group methods for
  quantum spin systems},\ }\href {https://doi.org/10.1080/14789940801912366}
  {\bibfield  {journal} {\bibinfo  {journal} {Advances in Physics}\ }\textbf
  {\bibinfo {volume} {57}},\ \bibinfo {pages} {143–224} (\bibinfo {year}
  {2008})}\BibitemShut {NoStop}%
\bibitem [{\citenamefont {Lehoucq}\ \emph {et~al.}(1998)\citenamefont
  {Lehoucq}, \citenamefont {Sorensen},\ and\ \citenamefont
  {Yang}}]{lehoucq1998arpack}%
  \BibitemOpen
  \bibfield  {author} {\bibinfo {author} {\bibfnamefont {R.~B.}\ \bibnamefont
  {Lehoucq}}, \bibinfo {author} {\bibfnamefont {D.~C.}\ \bibnamefont
  {Sorensen}},\ and\ \bibinfo {author} {\bibfnamefont {C.}~\bibnamefont
  {Yang}},\ }\href {https://epubs.siam.org/doi/pdf/10.1137/1.9780898719628.fm}
  {\emph {\bibinfo {title} {ARPACK users' guide: solution of large-scale
  eigenvalue problems with implicitly restarted Arnoldi methods}}}\ (\bibinfo
  {publisher} {SIAM},\ \bibinfo {year} {1998})\BibitemShut {NoStop}%
\bibitem [{\citenamefont {Calabrese}\ and\ \citenamefont
  {Cardy}(2004)}]{calabrese2004entanglement}%
  \BibitemOpen
  \bibfield  {author} {\bibinfo {author} {\bibfnamefont {P.}~\bibnamefont
  {Calabrese}}\ and\ \bibinfo {author} {\bibfnamefont {J.}~\bibnamefont
  {Cardy}},\ }\bibfield  {title} {\bibinfo {title} {Entanglement entropy and
  quantum field theory},\ }\href@noop {} {\bibfield  {journal} {\bibinfo
  {journal} {Journal of Statistical Mechanics: Theory and Experiment}\ }\textbf
  {\bibinfo {volume} {2004}},\ \bibinfo {pages} {P06002} (\bibinfo {year}
  {2004})}\BibitemShut {NoStop}%
\bibitem [{\citenamefont {Hovi}\ \emph {et~al.}(1996)\citenamefont {Hovi},
  \citenamefont {Aharony}, \citenamefont {Stauffer},\ and\ \citenamefont
  {Mandelbrot}}]{hovi1996gap}%
  \BibitemOpen
  \bibfield  {author} {\bibinfo {author} {\bibfnamefont {J.-P.}\ \bibnamefont
  {Hovi}}, \bibinfo {author} {\bibfnamefont {A.}~\bibnamefont {Aharony}},
  \bibinfo {author} {\bibfnamefont {D.}~\bibnamefont {Stauffer}},\ and\
  \bibinfo {author} {\bibfnamefont {B.~B.}\ \bibnamefont {Mandelbrot}},\
  }\bibfield  {title} {\bibinfo {title} {Gap independence and lacunarity in
  percolation clusters},\ }\href
  {https://journals.aps.org/prl/abstract/10.1103/PhysRevLett.77.877} {\bibfield
   {journal} {\bibinfo  {journal} {Physical review letters}\ }\textbf {\bibinfo
  {volume} {77}},\ \bibinfo {pages} {877} (\bibinfo {year} {1996})}\BibitemShut
  {NoStop}%
\bibitem [{\citenamefont {Ginelli}\ \emph {et~al.}(2005)\citenamefont
  {Ginelli}, \citenamefont {Hinrichsen}, \citenamefont {Livi}, \citenamefont
  {Mukamel},\ and\ \citenamefont {Politi}}]{ginelli2005directed}%
  \BibitemOpen
  \bibfield  {author} {\bibinfo {author} {\bibfnamefont {F.}~\bibnamefont
  {Ginelli}}, \bibinfo {author} {\bibfnamefont {H.}~\bibnamefont {Hinrichsen}},
  \bibinfo {author} {\bibfnamefont {R.}~\bibnamefont {Livi}}, \bibinfo {author}
  {\bibfnamefont {D.}~\bibnamefont {Mukamel}},\ and\ \bibinfo {author}
  {\bibfnamefont {A.}~\bibnamefont {Politi}},\ }\bibfield  {title} {\bibinfo
  {title} {Directed percolation with long-range interactions: Modeling
  nonequilibrium wetting},\ }\href
  {https://journals.aps.org/pre/abstract/10.1103/PhysRevE.71.026121} {\bibfield
   {journal} {\bibinfo  {journal} {Physical Review E}\ }\textbf {\bibinfo
  {volume} {71}},\ \bibinfo {pages} {026121} (\bibinfo {year}
  {2005})}\BibitemShut {NoStop}%
\bibitem [{\citenamefont {Gillespie}(1977)}]{gillespie1977exact}%
  \BibitemOpen
  \bibfield  {author} {\bibinfo {author} {\bibfnamefont {D.~T.}\ \bibnamefont
  {Gillespie}},\ }\bibfield  {title} {\bibinfo {title} {Exact stochastic
  simulation of coupled chemical reactions},\ }\href@noop {} {\bibfield
  {journal} {\bibinfo  {journal} {The journal of physical chemistry}\ }\textbf
  {\bibinfo {volume} {81}},\ \bibinfo {pages} {2340} (\bibinfo {year}
  {1977})}\BibitemShut {NoStop}%
\bibitem [{\citenamefont {Vidal}(2004)}]{vidal2004efficient}%
  \BibitemOpen
  \bibfield  {author} {\bibinfo {author} {\bibfnamefont {G.}~\bibnamefont
  {Vidal}},\ }\bibfield  {title} {\bibinfo {title} {Efficient simulation of
  one-dimensional quantum many-body systems},\ }\href@noop {} {\bibfield
  {journal} {\bibinfo  {journal} {Physical review letters}\ }\textbf {\bibinfo
  {volume} {93}},\ \bibinfo {pages} {040502} (\bibinfo {year}
  {2004})}\BibitemShut {NoStop}%
\end{thebibliography}%

\end{document}